\input pictex.tex   %% Dovrebbe essere in path (altrimenti vedi in Doc)

% $Id: dcpic.sty,v 1.24 2002/11/25 13:51:57 pedro Exp $
%% DC-PiCTeX
%% Realizado por Pedro Quaresma de Almeida, Coimbra 
%% 11/1990 (vers{\~a}o 1.0); 10/1991 (vers{\~a}o 1.1);
%%  9/1993 (vers{\~a}o 1.2);  3/1995 (vers{\~a}o 1.3);
%%  7/1996 (vers{\~a}o 2.1);
%%  5/2001 (vers{\~a}o 3.0); 11/2001 (vers{\~a}o 3.1);
%%  1/2002 (vers{\~a}o 3.2)
%%  5/2002 (versão 4.0)
\immediate\write10{Package DCpic 2002/05/16 v4.0}

\catcode`!=11 %  ***** THIS MUST NEVER BE OMITTED (Ver PiCTeX)

\newcount\aux%
\newcount\auxa%
\newcount\auxb%
\newcount\m%
\newcount\n%
\newcount\x%
\newcount\y%
\newcount\xl%
\newcount\yl%
\newcount\d%
\newcount\dnm%
\newcount\xa%
\newcount\xb%
\newcount\xmed%
\newcount\xc%
\newcount\xd%
\newcount\ya%
\newcount\yb%
\newcount\ymed%
\newcount\yc%
\newcount\yd
%% "variáveis globais"
\newcount\expansao%
\newcount\tipografo%       versão 4.0
\newcount\distanciaobjmor% versão 4.0
\newcount\tipoarco%        versão 4.0
%\newif\ifarredondada%        versão 4.0 (valor inicial "false")
\newif\ifpara%
%% version 3.2
\newbox\caixa%
\newbox\caixaaux%
\newif\ifnvazia%
\newif\ifvazia%
\newif\ifcompara%
\newif\ifdiferentes%
\newcount\xaux%
\newcount\yaux%
\newcount\guardaauxa%
\newcount\alt%
\newcount\larg%
\newcount\prof%
%% para os ajustes
\newcount\auxqx
\newcount\auxqy
\newif\ifajusta%
\newif\ifajustadist
\def\objPartida{}%
\def\objChegada{}%
\def\objNulo{}%

%% 
%% Stack specification
%%

%%
%% Emtpy stack
%%
\def\!vazia{:}

%%
%% Is Empty? : Stack -> Bool
%%
%% nvazia - True if Not Empy
%% vazia  - True if Empty
\def\!pilhanvazia#1{\let\arg=#1%
\if:\arg\ \nvaziafalse\vaziatrue \else \nvaziatrue\vaziafalse\fi}

%%
%% Push : Elems x Stack -> Stack
%%
\def\!coloca#1#2{\edef\pilha{#1.#2}}

%%
%% Top : Stack -> Elems
%%
%% the empty stack is not taken care
%% the element is "kept" ("guardado") 
\def\!guarda(#1)(#2,#3)(#4,#5,#6){\def\id{#1}%
\xaux=#2%
\yaux=#3%
\alt=#4%
\larg=#5%
\prof=#6%
}

\def\!topaux#1.#2:{\!guarda#1}
\def\!topo#1{\expandafter\!topaux#1}

%%
%% Pop : Stack -> Stack
%%
%% the empty stack is not taken care
\def\!popaux#1.#2:{\def\pilha{#2:}}
\def\!retira#1{\expandafter\!popaux#1}

%%
%% Compares words : Word x Word -> Bool
%%
%% compara - True if equal
%% diferentes - True if not equal
\def\!comparaaux#1#2{\let\argA=#1\let\argB=#2%
\ifx\argA\argB\comparatrue\diferentesfalse\else\comparafalse\diferentestrue\fi}

\def\!compara#1#2{\!comparaaux{#1}{#2}}

%%Comando Interno
%% Valor absoluto (absolute value)
%% \absoluto{n}{absn}
%% entrada
%%  n - natural
%% sa{\'\i}da
%%  absn - o valor absoluto de n
\def\!absoluto#1#2{\n=#1%
  \ifnum \n > 0
    #2=\n
  \else
    \multiply \n by -1
    #2=\n
  \fi}

%% Name definitions for edge types and directions

\def\dasharrow{1}
\def\solidline{2}

%% Name definitions for edge label placement

\def\atleft{1}
%% Tip direction for curved edges

\def\pright{2}

%% Type of graph
\def\commdiag{0}

%% Posicionamento da etiquetas nos grafos

%%Comando Interno
%% Ajusta a dist{\^a}ncia entre as setas e os objectos em fun{\c c}{\~a}o das
%% dimens{\~o}es destes {\'u}ltimos
%% \ajusta{x}{xl}{y}{yl}{d}{Objecto}
%% entrada
%%  (x,y) e (xl,yl), coordenadas dos pontos de {\'\i}nicio e fim da seta
%%  d, dist{\^a}ncia especificada pelo utilizador ou 10 (valor por
%%  omiss{\~a}o), Objecto d{\'a}-nos a refer{\^e}ncia do objecto ao qual se est{\'a} a
%%  efectuar o ajuste.
%% sa{\'\i}da
%%  d, dist{\^a}ncia alterada.
%% 
%% A dist{\^a}ncia alterada {\'e} o maior valor entre 10 e as dimens{\~o}es
%% apropriadas da caixa que cont{\^e}m o objecto. 
%% Se o utilizador especificar um valor essa especifica{\c c}{\~a}o
%% n{\~a}o {\'e} alterada.
%%
%% Se a seta {\'e} horizontal usa-se o valor da largura
%% Se a seta {\'e} vertical usa-se:
%%  o valor da altura se a seta est{\'a} no 1o ou 2o quadrante
%%  o valor da profundidade se a seta est{\'a} no 3o ou 4o quadrante
%% Se a seta {\'e} {\'o}bliqua vai-se escolher o valor conforme:
%%  de 315 a  45 graus usa-se a largura
%%  de  45 a 135 graus usa-se a altura
%%  de 135 a 225 graus usa-se a largura
%%  de 225 a 315 graus usa-se a profundidade
\def\!ajusta#1#2#3#4#5#6{\aux=#5%
  \let\auxobj=#6%
  \ifcase \tipografo    % diagramas comutativos
    \ifnum\number\aux=10 
      \ajustadisttrue % se o valor é o valor por omissão ajusta
    \else
      \ajustadistfalse  % caso contrário não ajusta
    \fi
  \else  % grafos (dirigidos, não dirigidos, com molduras)
   \ajustadistfalse
%  \or  % grafos não dirigidos
%   \ajustadistfalse
%  \else % grafos dirigidos com molduras circulares nos objectos
%    \ifnum\number\aux=8 
%      \ajustadisttrue  % se o valor é o valor por omissão ajusta
%    \else
%      \ajustadistfalse % caso contrário não ajusta
%    \fi
  \fi
  \ifajustadist
%  \tiny Vou ajustar %%%%%%%%%%%%%%%%%%%%%%%%%%%%%%
%  \ifnum\number\aux=10% verificar se s{\~a}o os valores por omiss{\~a}o
   \let\pilhaaux=\pilha%
   \loop%
     \!topo{\pilha}%
     \!retira{\pilha}%
     \!compara{\id}{\auxobj}%
     \ifcompara\nvaziafalse \else\!pilhanvazia\pilha \fi%
     \ifnvazia%
   \repeat%
%% rep{\~o}e os valores na pilha
   \let\pilha=\pilhaaux%
   \ifvazia%
    \ifdiferentes%
%%
%% N{\~a}o {\'e} poss{\'\i}vel efectuar o ajuste dado o utilizador n{\~a}o ter
%% especificado uma etiqueta para o objecto em quest{\~a}o. {\'E} dado o
%% valor de 10, igual ao valor por omiss{\~a}o.
%%
     \larg=1310720% n{\~a}o faz o ajuste
     \prof=655360%
     \alt=655360%
    \fi%
   \fi%
   \divide\larg by 131072
   \divide\prof by 65536
   \divide\alt by 65536
   \ifnum\number\y=\number\yl
%% Caso 1 -- seta horizontal
%%
%% divide-se por 131072 para se obter metade da largura da caixa em
%% pontos (pt), isto dado que o texto est{\'a} centrado na caixa. Soma-se
%% mais tr{\^e}s, que constitue um ajuste imp{\'\i}rico.
    \advance\larg by 3
    \ifnum\number\larg>\aux
     #5=\larg
    \fi
   \else
    \ifnum\number\x=\number\xl
     \ifnum\number\yl>\number\y
%% Caso 2.1 -- seta vertical de cima para baixa
%%
      \ifnum\number\alt>\aux
       #5=\alt
      \fi
     \else
%% Caso 2.2 -- seta vertical de baixo para cima
%%
%% divide-se por 65536 para se obter a altura da caixa em pt. O ajuste
%% de 5 foi obtido imp{\'\i}ricamente
      \advance\prof by 5
      \ifnum\number\prof>\aux
       #5=\prof
      \fi
     \fi
    \else
%% Caso 3 -- seta obl{\'\i}qua 
%% Caso 3.1 de 315o a  45o; |x-xl|>|y-yl| e
%% Caso 3.3 de 135o a 225o; |x-xl|>|y-yl|; Largura
     \auxqx=\x
     \advance\auxqx by -\xl
     \!absoluto{\auxqx}{\auxqx}%
     \auxqy=\y
     \advance\auxqy by -\yl
     \!absoluto{\auxqy}{\auxqy}%
     \ifnum\auxqx>\auxqy
      \ifnum\larg<10
       \larg=10
      \fi
      \advance\larg by 3
      #5=\larg
     \else
%% Caso 3.2 de  45o a 135o; |x-xl|<|y-yl| e y>0; Largura
      \ifnum\yl>\y
       \ifnum\larg<10
        \larg=10
       \fi
      \advance\alt by 6
       #5=\alt
      \else
%% Caso 3.4 de 225o a 315o; |x-xl|<|y-yl| e y<0; Profundidade
      \advance\prof by 11
       #5=\prof
      \fi
     \fi
    \fi
   \fi
\fi} % o ramo "else" {\'e} omisso

%%Comando Interno
%% C{\'a}lculo da Raiz Quadrada
%% raiz{n}{m}
%% entrada
%%   n - natural
%% sa{\'\i}da
%%   n - natural
%%   m - maior natural contido na raiz quadrada de n
\def\!raiz#1#2{\n=#1%
  \m=1%
  \loop
    \aux=\m%
    \advance \aux by 1%
    \multiply \aux by \aux%
    \ifnum \aux < \n%
      \advance \m by 1%
      \paratrue%
    \else\ifnum \aux=\n%
      \advance \m by 1%
      \paratrue%
       \else\parafalse%
       \fi
    \fi
  \ifpara%
  \repeat
#2=\m}

%%Comando Interno
%% Calcula os pontos de 
%%       come{\c c}o da "seta"
%%       fim da "seta"
%%   coloca{\c c}{\~a}o do s{\'\i}mbolo
%% 
%% ucoord{x1}{x2}{x3}{x4}{x5}{x6}{+|- 1}
%% entrada
%%   x1,x2,x3,x4,x5
%% sa{\'\i}da
%%   x6
%%  
%%              x2 - x1
%%  x6 = x3 +|- ------- x4
%%                 x5
\def\!ucoord#1#2#3#4#5#6#7{\aux=#2%
  \advance \aux by -#1%
  \multiply \aux by #4%
  \divide \aux by #5%
  \ifnum #7 = -1 \multiply \aux by -1 \fi%
  \advance \aux by #3%
#6=\aux}

%%Comando Interno 
%% C{\'a}lculo do Quadrado da Dist{\^a}ncia Euclidiana entre dois pontos 
%% quadrado{n}{m}{l}
%% entrada
%%   n - natural
%%   m - natural
%% sa{\'\i}da
%%   l = (n-m)*(n-m)
\def\!quadrado#1#2#3{\aux=#1%
  \advance \aux by -#2%
  \multiply \aux by \aux%
#3=\aux}

%%Comando Interno
%% C{\'a}lculo auxiliar para determinar a dist{\^a}ncia entre o nome do
%% morfismo e a seta.
%% entrada
%%     (x,y), (x',y') e o nome do morfismo
%% sa{\'\i}da
%%     dnm - dist{\^a}ncia do nome ao morfismo respectivo devidamente
%%     compensada pelo tamanho do objecto
%% Observa{\c c}{\~o}es
%%     A compensa{\c c}{\~a}o s{\'o} est{\'a} a ser feita para setas
%%     horizontais e verticais. As obl{\'\i}quas s{\~a}o tratadas de
%%     outra forma.
%% algoritmo
%%  caixa0 <- nome do morfismo
%%  se x-xl = 0 entao                   {recta vertical}
%%     aux <- largura da caixa0
%%     dnm <- convers{\~a}o-sp-pt(aux)/2+3
%%  sen{\~a}o                               {recta n{\~a}o vertical}
%%     se y-yl = 0 entao                {recta horizontal}
%%        aux <- altura+profundidade da caixa0
%%        dnm <- convers{\~a}o-sp-pt(aux)/2+3
%%     sen{\~a}o                            {recta obl{\'\i}qua}
%%        dnm <- 3
%%     fimse
%%  fimse
%% fimalgoritmo
\def\!distnomemor#1#2#3#4#5#6{\setbox0=\hbox{#5}%
  \aux=#1
  \advance \aux by -#3
  \ifnum \aux=0
     \aux=\wd0 \divide \aux by 131072
     \advance \aux by 3
     #6=\aux
  \else
     \aux=#2
     \advance \aux by -#4
     \ifnum \aux=0
        \aux=\ht0 \advance \aux by \dp0 \divide \aux by 131072
        \advance \aux by 3
        #6=\aux%
     \else
     #6=3
     \fi
   \fi
}

%%
%% O ambiente "begindc...enddc"
%%
\def\begindc#1{\!ifnextchar[{\!begindc{#1}}{\!begindc{#1}[30]}}
\def\!begindc#1[#2]{\beginpicture 
  \let\pilha=\!vazia
  \setcoordinatesystem units <1pt,1pt>
  \expansao=#2
  \ifcase #1
    \distanciaobjmor=10
    \tipoarco=0         % seta
    \tipografo=0        % diagrama comutativo
  \or
    \distanciaobjmor=2
    \tipoarco=0         % seta 
    \tipografo=1        % grafo orientado
  \or
    \distanciaobjmor=1
    \tipoarco=2         % linha
    \tipografo=2        % grafo não orientado
  \or
    \distanciaobjmor=8
    \tipoarco=0         % seta 
    \tipografo=3        % grafo orientado
%    \arredondadotrue    % objectos com molduras circulares
  \or
    \distanciaobjmor=8
    \tipoarco=2         % linha
    \tipografo=4        % grafo não orientado
%    \arredondadotrue    % objectos com molduras circulares
  \fi}

\def\enddc{\endpicture}

%%
%% Comando para construir a "seta" entre dois objectos
%%
%% Os pontos definidores da seta e da etiqueta respectiva s{\~a}o:
%% 
%%                (xd,yd)
%%                   o
%%                   |
%%  o------o---------o---------o------o
%%(x,y) (xa,ya)   (xm,ym)   (xb,yb)(xl,yl)
%%
\def\mor{%
  \!ifnextchar({\!morxy}{\!morObjA}}
\def\!morxy(#1,#2){%
  \!ifnextchar({\!morxyl{#1}{#2}}{\!morObjB{#1}{#2}}}
\def\!morxyl#1#2(#3,#4){%
  \!ifnextchar[{\!mora{#1}{#2}{#3}{#4}}{\!mora{#1}{#2}{#3}{#4}[\number\distanciaobjmor,\number\distanciaobjmor]}}%
\def\!morObjA#1{%
 \let\pilhaaux=\pilha%
 \def\objPartida{#1}%
 \loop%
    \!topo\pilha%
    \!retira\pilha%
    \!compara{\id}{\objPartida}%
    \ifcompara \nvaziafalse \else \!pilhanvazia\pilha \fi%
   \ifnvazia%
 \repeat%
 \ifvazia%
  \ifdiferentes%
%%
%% Mensagem de erro e atribui{\c c}{\~a}o de valores fict{\'\i}cios aos 
%% argumentos dos comandos que se seguem.
%%
   Error: Incorrect label specification%
   \xaux=1%
   \yaux=1%
  \fi%
 \fi% 
 \let\pilha=\pilhaaux%
 \!ifnextchar({\!morxyl{\number\xaux}{\number\yaux}}{\!morObjB{\number\xaux}{\number\yaux}}}
\def\!morObjB#1#2#3{%
  \x=#1
  \y=#2
 \def\objChegada{#3}%
 \let\pilhaaux=\pilha%
 \loop
    \!topo\pilha %
    \!retira\pilha%
    \!compara{\id}{\objChegada}%
    \ifcompara \nvaziafalse \else \!pilhanvazia\pilha \fi
   \ifnvazia
 \repeat
 \ifvazia
  \ifdiferentes%
%%
%% Mensagem de erro e atribui{\c c}{\~a}o de valores fict{\'\i}cios aos 
%% argumentos dos comandos que se seguem.
%%
   Error: Incorrect label specification
   \xaux=\x%
   \advance\xaux by \x%
   \yaux=\y%
   \advance\yaux by \y%
  \fi
 \fi
 \let\pilha=\pilhaaux
 \!ifnextchar[{\!mora{\number\x}{\number\y}{\number\xaux}{\number\yaux}}{\!mora{\number\x}{\number\y}{\number\xaux}{\number\yaux}[\number\distanciaobjmor,\number\distanciaobjmor]}}
\def\!mora#1#2#3#4[#5,#6]#7{%
  \!ifnextchar[{\!morb{#1}{#2}{#3}{#4}{#5}{#6}{#7}}{\!morb{#1}{#2}{#3}{#4}{#5}{#6}{#7}[1,\number\tipoarco] }}
\def\!morb#1#2#3#4#5#6#7[#8,#9]{\x=#1%
  \y=#2%
  \xl=#3%
  \yl=#4%
  \multiply \x by \expansao%
  \multiply \y by \expansao%
  \multiply \xl by \expansao%
  \multiply \yl by \expansao%
%%
%% calcular a dist{\^a}ncia Euclidiana entre dois pontos
%% d = \sqrt((x-xl)^2+(y-yl)^2)
%%
  \!quadrado{\number\x}{\number\xl}{\auxa}%
  \!quadrado{\number\y}{\number\yl}{\auxb}%
  \d=\auxa%
  \advance \d by \auxb%
  \!raiz{\d}{\d}%
%%
%% o ponto (xa,ya) est{\'a} {\`a} dist{\^a}ncia #5 (valor por omiss{\~a}o 10) do ponto
%% (x,y)
%%
%% como existem dois pontos em considera{\c c}{\~a}o, o ponto de partida e o
%% ponto de chegada, vai sei necess{\'a}rio recuperar de novo os seus
%% valores por pesquisa na pilha
  \auxa=#5
  \!compara{\objNulo}{\objPartida}%
  \ifdiferentes% S{\'o} vai fazer o ajuste quando {\'e} necess{\'a}rio
   \!ajusta{\x}{\xl}{\y}{\yl}{\auxa}{\objPartida}%
   \ajustatrue
   \def\objPartida{}% re-inicializar o valor do Objecto de Partida
  \fi
%% vai guardar o valor de auxa (ap{\'o}s ajuste) para ser usado no caso
%% dos morfismos de injec{\c c}{\~a}o.
  \guardaauxa=\auxa
  \!ucoord{\number\x}{\number\xl}{\number\x}{\auxa}{\number\d}{\xa}{1}%
  \!ucoord{\number\y}{\number\yl}{\number\y}{\auxa}{\number\d}{\ya}{1}%
%% auxa vai ter o valor da dist{\^a}ncia entre os objectos menos a
%% dist{\^a}ncia da seta ao objecto (10 por omiss{\~a}o)
  \auxa=\d%
%%
%% o ponto (xb,yb) est{\'a} {\`a} dist{\^a}ncia #6 (valor por omiss{\~a}o 10) do ponto
%% (xl,yl)
%%
  \auxb=#6
  \!compara{\objNulo}{\objChegada}%
  \ifdiferentes% S{\'o} vai fazer o ajuste quando {\'e} necess{\'a}rio
%   Vou ajustar
   \!ajusta{\x}{\xl}{\y}{\yl}{\auxb}{\objChegada}%
   \def\objChegada{}% re-inicializar o valor do Objecto de Chegada
  \fi
  \advance \auxa by -\auxb%
  \!ucoord{\number\x}{\number\xl}{\number\x}{\number\auxa}{\number\d}{\xb}{1}%
  \!ucoord{\number\y}{\number\yl}{\number\y}{\number\auxa}{\number\d}{\yb}{1}%
  \xmed=\xa%
  \advance \xmed by \xb%
  \divide \xmed by 2
  \ymed=\ya%
  \advance \ymed by \yb%
  \divide \ymed by 2
  \!distnomemor{\number\x}{\number\y}{\number\xl}{\number\yl}{#7}{\dnm}%
  \!ucoord{\number\y}{\number\yl}{\number\xmed}{\number\dnm}{\number\d}{\xc}{-#8}% 
  \!ucoord{\number\x}{\number\xl}{\number\ymed}{\number\dnm}{\number\d}{\yc}{#8}%
\ifcase #9  % seta s{\'o}lida
  \arrow <4pt> [.2,1.1] from {\xa} {\ya} to {\xb} {\yb}
\or  % seta a tracejado
  \setdashes
  \arrow <4pt> [.2,1.1] from {\xa} {\ya} to {\xb} {\yb}
  \setsolid
\or  % linha s{\'o}lida
  \setlinear
  \plot {\xa} {\ya}  {\xb} {\yb} /
\or  % seta de injec{\c c}{\~a}o
%% C{\'a}lculos auxiliares
%%
%% 3 valor para o raio do "rabo" da "seta"
%%
%% repor o valor de auxa
  \auxa=\guardaauxa
%% dar a compensa{\c c}{\~a}o para o "rabo"
  \advance \auxa by 3%
%%
%% IMPORTANTE os valores de xa e ya v{\~a}o ser alterados
%%
 \!ucoord{\number\x}{\number\xl}{\number\x}{\number\auxa}{\number\d}{\xa}{1}%
 \!ucoord{\number\y}{\number\yl}{\number\y}{\number\auxa}{\number\d}{\ya}{1}%
 \!ucoord{\number\y}{\number\yl}{\number\xa}{3}{\number\d}{\xd}{-1}%
 \!ucoord{\number\x}{\number\xl}{\number\ya}{3}{\number\d}{\yd}{1}%
%% Constru{\c c}{\~a}o da "seta"
  \arrow <4pt> [.2,1.1] from {\xa} {\ya} to {\xb} {\yb}
%% e do seu "rabo"
  \circulararc -180 degrees from {\xa} {\ya} center at {\xd} {\yd}
\or  % seta de aplica{\c c}{\~a}o ("|-->")
  \auxa=3% valor para o meio-segmento do "rabo" da "seta"
%% c{\'a}lculo dos pontos (xmed,ymed) e (xd,yd) para o segmento de recta que
%% define o "rabo" da seta
 \!ucoord{\number\y}{\number\yl}{\number\xa}{\number\auxa}{\number\d}{\xmed}{-1}%
 \!ucoord{\number\x}{\number\xl}{\number\ya}{\number\auxa}{\number\d}{\ymed}{1}%
 \!ucoord{\number\y}{\number\yl}{\number\xa}{\number\auxa}{\number\d}{\xd}{1}%
 \!ucoord{\number\x}{\number\xl}{\number\ya}{\number\auxa}{\number\d}{\yd}{-1}%
%% Constru{\c c}{\~a}o da "seta"
  \arrow <4pt> [.2,1.1] from {\xa} {\ya} to {\xb} {\yb}
%% e do seu "rabo"
  \setlinear
  \plot {\xmed} {\ymed}  {\xd} {\yd} /
\fi
%% Coloca{\c c}{\~a}o do nome do morfismo.
%% Se os morfismos s{\~a}o horizontais ou verticais constro{\'\i} uma caixa
%% centrada no ponto pr{\'e}viamente calculado. Se as setas s{\~a}o
%% obl{\'\i}quas coloca a caixa de forma a n{\~a}o colidir com o morfismo 
%% tendo em aten{\c c}{\~a}o o quadrante assim como a posi{\c c}{\~a}o
%% relativa do morfismo e do respectivo nome.
\auxa=\xl
\advance \auxa by -\x%
\ifnum \auxa=0 
  \put {#7} at {\xc} {\yc}
\else
  \auxb=\yl
  \advance \auxb by -\y%
  \ifnum \auxb=0 \put {#7} at {\xc} {\yc}
  \else 
    \ifnum \auxa > 0 
      \ifnum \auxb > 0
        \ifnum #8=1
          \put {#7} [rb] at {\xc} {\yc}
        \else 
          \put {#7} [lt] at {\xc} {\yc}
        \fi
      \else
        \ifnum #8=1
          \put {#7} [lb] at {\xc} {\yc}
        \else 
          \put {#7} [rt] at {\xc} {\yc}
        \fi
      \fi
    \else
      \ifnum \auxb > 0 
        \ifnum #8=1
          \put {#7} [rt] at {\xc} {\yc}
        \else 
          \put {#7} [lb] at {\xc} {\yc}
        \fi
      \else
        \ifnum #8=1
          \put {#7} [lt] at {\xc} {\yc}
        \else 
          \put {#7} [rb] at {\xc} {\yc}
        \fi
      \fi
    \fi
  \fi
\fi
}

%%
%% Comando para construir a "seta" curvilinea entre dois objectos
%%
%% \cmor(<lista de pontos (n. impar)>){<etiqueta>}
%%
%% Em primeiro lugar {\'e} necess{\'a}rio modificar o comando plot de forma a
%% que a sintaxe de utiliza{\c c}{\~a}o do novo comando seja coerente com a
%% sintaxe dos restantes comandos
%%
\def\modifplot(#1{\!modifqcurve #1}
\def\!modifqcurve(#1,#2){\x=#1%
  \y=#2%
  \multiply \x by \expansao%
  \multiply \y by \expansao%
  \!start (\x,\y)
  \!modifQjoin}
\def\!modifQjoin(#1,#2)(#3,#4){\x=#1%
  \y=#2%
  \xl=#3%
  \yl=#4%
  \multiply \x by \expansao%
  \multiply \y by \expansao%
  \multiply \xl by \expansao%
  \multiply \yl by \expansao%
  \!qjoin (\x,\y) (\xl,\yl)             % \!qjoin  is defined in QUADRATIC
  \!ifnextchar){\!fim}{\!modifQjoin}}
\def\!fim){\ignorespaces}

%%
%% O comando para desenhar a seta vai receber a lista de pontos da qual
%% retira o {\'u}ltimo par de pontos, dependente da escolha dada pelo
%% utilizador a seta vai ser desenhada para cima, para baixo, para a
%% direita ou para a esquerda
%%
\def\setaxy(#1{\!pontosxy #1}
\def\!pontosxy(#1,#2){%
  \!maispontosxy}
\def\!maispontosxy(#1,#2)(#3,#4){%
  \!ifnextchar){\!fimxy#3,#4}{\!maispontosxy}}
\def\!fimxy#1,#2){\x=#1%
  \y=#2
  \multiply \x by \expansao
  \multiply \y by \expansao
  \xl=\x%
  \yl=\y%
  \aux=1%
  \multiply \aux by \auxa%
  \advance\xl by \aux%
  \aux=1%
  \multiply \aux by \auxb%
  \advance\yl by \aux%
  \arrow <4pt> [.2,1.1] from {\x} {\y} to {\xl} {\yl}}

%%
%% Temos agora a defini{\c c}{\~a}o do comando "cmor"
%%
\def\cmor#1 #2(#3,#4)#5{%
  \!ifnextchar[{\!cmora{#1}{#2}{#3}{#4}{#5}}{\!cmora{#1}{#2}{#3}{#4}{#5}[0] }}
\def\!cmora#1#2#3#4#5[#6]{%
  \ifcase #2% para cima "\pup" (pointing up)
      \auxa=0% x mant{\^e}m-se
      \auxb=1% o y "sobe" 
    \or% para baixo "\pdown" (pointing down)
      \auxa=0% x mant{\^e}m-se
      \auxb=-1% o y "desce" 
    \or% para a direita "\pright" (pointing right)
      \auxa=1% o x move-se para a direita
      \auxb=0% o y mant{\^e}m-se
    \or% para a esquerda "\pleft" (pointing left)
      \auxa=-1% o x move-se para a esquerda
      \auxb=0% o y mant{\^e}m-se
    \fi  % constru{\c c}{\~a}o do arco
  \ifcase #6  % arco (com seta) s{\'o}lido
    \modifplot#1% Desenhar o arco
    % constru{\c c}{\~a}o da seta
    \setaxy#1
  \or  % arco (com seta) a tracejado
    \setdashes
    \modifplot#1% Desenhar o arco
    \setaxy#1
    \setsolid
  \or  % arco sem seta
    \modifplot#1% Desenhar o arco
  \fi  % seta de injec{\c c}{\~a}o
%% coloca{\c c}{\~a}o da etiqueta do morfismo
  \x=#3%  
  \y=#4%
  \multiply \x by \expansao%
  \multiply \y by \expansao%
  \put {#5} at {\x} {\y}}

%%
%% Comando para construir os Objectos
%%  \obj(x,y){<texto>}[<etiqueta>]
%% 
\def\obj(#1,#2){%
  \!ifnextchar[{\!obja{#1}{#2}}{\!obja{#1}{#2}[Nulo]}}
\def\!obja#1#2[#3]#4{%
  \!ifnextchar[{\!objb{#1}{#2}{#3}{#4}}{\!objb{#1}{#2}{#3}{#4}[1]}}
\def\!objb#1#2#3#4[#5]{%
  \x=#1%
  \y=#2%
  \def\!pinta{\normalsize$\bullet$}% para definir o tamanho normal das pintas
  \def\!nulo{Nulo}%
  \def\!arg{#3}%
  \!compara{\!arg}{\!nulo}%
  \ifcompara\def\!arg{#4}\fi%
  \multiply \x by \expansao%
  \multiply \y by \expansao%
  \setbox\caixa=\hbox{#4}%
  \!coloca{(\!arg)(#1,#2)(\number\ht\caixa,\number\wd\caixa,\number\dp\caixa)}{\pilha}%
  \auxa=\wd\caixa \divide \auxa by 131072 
  \advance \auxa by 5
  \auxb=\ht\caixa
  \advance \auxb by \number\dp\caixa
  \divide \auxb by 131072 
  \advance \auxb by 5
%(\number\auxa,
%\number\auxb)
%  \aux=\ht\caixa \divide \auxa by 131072 
% \advance \auxa by 5 
%  \auxb=\dp\caixa \divide \auxb by 131072 
%  \advance \auxb by 8
  \ifcase \tipografo    % diagramas comutativos
    \put{#4} at {\x} {\y}
  \or                   % grafos dirigidos
    \ifcase #5 % c=0
      \put{#4} at {\x} {\y}
    \or        % n=1
      \put{\!pinta} at {\x} {\y}
      \advance \y by \number\auxb  % height+depth+5
      \put{#4} at {\x} {\y}
    \or        % ne=2
      \put{\!pinta} at {\x} {\y}
      \advance \auxa by -2  % para fazer o ajuste (imperfeito)
      \advance \auxb by -2  % ao raio da circunferência de centro (x,y)
      \advance \x by \number\auxa  % width+5
      \advance \y by \number\auxb  % height+depth+5
      \put{#4} at {\x} {\y}   
    \or        % e=3
      \put{\!pinta} at {\x} {\y}
      \advance \x by \number\auxa  % width+5
      \put{#4} at {\x} {\y}   
    \or        % se=4
      \put{\!pinta} at {\x} {\y}
      \advance \auxa by -2  % para fazer o ajuste (imperfeito)
      \advance \auxb by -2  % ao raio da circunferência de centro (x,y)
      \advance \x by \number\auxa  % width+5
      \advance \y by -\number\auxb  % height+depth+5
      \put{#4} at {\x} {\y}   
    \or        % s=5
      \put{\!pinta} at {\x} {\y}
      \advance \y by -\number\auxb  % height+depth+5
      \put{#4} at {\x} {\y}   
    \or        % sw=6
      \put{\!pinta} at {\x} {\y}
      \advance \auxa by -2  % para fazer o ajuste (imperfeito)
      \advance \auxb by -2  % ao raio da circunferência de centro (x,y)
      \advance \x by -\number\auxa  % width+5
      \advance \y by -\number\auxb  % height+depth+5
      \put{#4} at {\x} {\y}   
    \or        % w=7
      \put{\!pinta} at {\x} {\y}
      \advance \x by -\number\auxa  % width+5
      \put{#4} at {\x} {\y}   
    \or        % nw=8
      \put{\!pinta} at {\x} {\y}
      \advance \auxa by -2  % para fazer o ajuste (imperfeito)
      \advance \auxb by -2  % ao raio da circunferência de centro (x,y)
      \advance \x by -\number\auxa  % width+5
      \advance \y by \number\auxb  % height+depth+5
      \put{#4} at {\x} {\y}   
    \fi
  \or                   % grafos não dirigidos
    \ifcase #5 % c=0
      \put{#4} at {\x} {\y}
    \or        % n=1
      \put{\!pinta} at {\x} {\y}
      \advance \y by \number\auxb  % height+depth+5
      \put{#4} at {\x} {\y}
    \or        % ne=2
      \put{\!pinta} at {\x} {\y}
      \advance \auxa by -2  % para fazer o ajuste (imperfeito)
      \advance \auxb by -2  % ao raio da circunferência de centro (x,y)
      \advance \x by \number\auxa  % width+5
      \advance \y by \number\auxb  % height+depth+5
      \put{#4} at {\x} {\y}   
    \or        % e=3
      \put{\!pinta} at {\x} {\y}
      \advance \x by \number\auxa  % width+5
      \put{#4} at {\x} {\y}   
    \or        % se=4
      \put{\!pinta} at {\x} {\y}
      \advance \auxa by -2  % ver acima
      \advance \auxb by -2
      \advance \x by \number\auxa  % width+5
      \advance \y by -\number\auxb % height+depth+5
      \put{#4} at {\x} {\y}   
    \or        % s=5
      \put{\!pinta} at {\x} {\y}
      \advance \y by -\number\auxb % height+depth+5
      \put{#4} at {\x} {\y}   
    \or        % sw=6
      \put{\!pinta} at {\x} {\y}
      \advance \auxa by -2  % ver acima
      \advance \auxb by -2
      \advance \x by -\number\auxa % width+5
      \advance \y by -\number\auxb % height+depth+5
      \put{#4} at {\x} {\y}   
    \or        % w=7
      \put{\!pinta} at {\x} {\y}
      \advance \x by -\number\auxa % width+5
      \put{#4} at {\x} {\y}   
    \or        % nw=8
      \put{\!pinta} at {\x} {\y}
      \advance \auxa by -2  % ver acima
      \advance \auxb by -2
      \advance \x by -\number\auxa % width+5
      \advance \y by \number\auxb  % height+depth+5
      \put{#4} at {\x} {\y}   
    \fi
%  \or % grafos dirigidos com molduras circulares nos objectos
%    \advance \auxa by 4
%    \put{\circle{\auxa}} [Bl] at {\x} {\y}
%    \put{#4} at {\x} {\y}
%  \or % grafos não dirigidos com molduras circulares nos objectos
   \else % grafos com molduras circulares nos objectos
     \ifnum\auxa<\auxb % determina a maior das dimensões
       \aux=\auxb
     \else
       \aux=\auxa
     \fi
% se a largura da caixa é menor do que 1em então o tamanho 
% tamanho é ajustado para esse valor mínimo
     \ifdim\wd\caixa<1em
       \dimen99 = 1em
       \aux=\dimen99 \divide \aux by 131072 
       \advance \aux by 5
     \fi
     \advance\aux by -2 %folga entre o objecto e a moldura
     \multiply\aux by 2 % 
     \ifnum\aux<30
       \put{\circle{\aux}} [Bl] at {\x} {\y}
     \else
       \multiply\auxa by 2
       \multiply\auxb by 2
       \put{\oval(\auxa,\auxb)} [Bl] at {\x} {\y}
     \fi
     \put{#4} at {\x} {\y}
   \fi   
}

\catcode`!=12 %  *****  THIS MUST NEVER BE OMITTED (Ver PiCTeX)

  \input miniltx
  \def\Gin@driver{pdftex.def}
  \input color.sty
  \input graphicx.sty
  \resetatcatcode

%\input graphicx.tex
%
%    Bundle of my macros.    Version 1.2.0.beta
%    The best use is to paste all of them into the papers
%     1/8/2005
%

%
%    Fonts.    Version 1.2.0.beta
%    The best use is to paste all of them into the papers
%     1/8/2005
%
%
% History:
%	3 Agosto 2005: ChernSimons.tex
%
%%%%%%%%%%%%%%%%%%%%%%%%%%%%

%%%%%%%%%%%%%%%%%%%%%%%%%%%%%%%%%%%%%%%%%%%%%%%%%%%%
%% Font Types	%%%%%%%%%%%%%%%%%%%%%%%%%%%%%%%%%%%%%%%%%%%%
%%%%%%%%%%%%%%%%%%%%%%%%%%%%%%%%%%%%%%%%%%%%%%%%%%%%%
\def\Serif{cmr}
\def\SerifBold{cmbx}
\def\SerifItalics{cmti}
\def\SerifSlanted{cmsl}
\def\SerifBoldItalics{cmbxti}
\def\SansSerif{cmss}
\def\SansSerifBold{cmssbx}
\def\SansSerifItalics{cmssi}
\def\SansSerifSlanted{cmssi}%%
\def\Math{cmmi}
\def\Symbols{cmsy}
\def\MathBold{cmmib}
\def\MoreSymbols{cmex}
\def\Typewriter{cmtt}
\def\Gothic{eufm}
\def\Double{msbm}
\def\Relazioni{msam}

%% Font Declarations	
%\font\tenbg=cmmib10%
%\def\bg{\tenbg}%
%%%%%%%%%%%%%%%%%%%%%%%%%%%%%%%%%%%%%%%%%%%%%%%%%%%%%%
%%%	5		%%%%%%%%%%%%%%%%%%%%%%%%%%%%%%%%%%%%%%%%%%%%
%%%	%%%%%%%%%%%%%%%%%%%%%%
= 			\Serif10 			at 5pt
= 		\SerifBold10 		at 5pt
= 	\SerifItalics10 	at 5pt
=		\SerifSlanted10 	at 5pt
=	\SerifBoldItalics10	at 5pt
= 		\SansSerif10 		at 5pt
=	\SansSerifBold10	at 5pt
=	\SansSerifItalics10	at 5pt
=	\SansSerifSlanted10	at 5pt
=				\Math10				at 5pt
=			\MathBold10			at 5pt
=			\Symbols10			at 5pt
=		\MoreSymbols10		at 5pt
=		\Typewriter10		at 5pt
=			\Gothic10			at 5pt
=			\Double10			at 5pt

%%%	7		%%%%%%%%%%%%%%%%%%%%%%%%%%%%%%%%%%%%%%%%%%%
%%%	%%%%%%%%%%%%%%%%%%%%%%%
= 			\Serif10 			at 7pt
= 		\SerifBold10 		at 7pt
= 	\SerifItalics10 	at 7pt
=	\SerifSlanted10 	at 7pt
=\SerifBoldItalics10	at 7pt
= 		\SansSerif10 		at 7pt
= 	\SansSerifBold10 	at 7pt
=\SansSerifItalics10	at 7pt
=\SansSerifSlanted10	at 7pt
=			\Math10				at 7pt
=		\MathBold10			at 7pt
=			\Symbols10			at 7pt
=		\MoreSymbols10		at 7pt
=		\Typewriter10		at 7pt
=			\Gothic10			at 7pt
=			\Double10			at 7pt

%%%	8		%%%%%%%%%%%%%%%%%%%%%%%%%%%%%%%%%%%%%%%%%
%%%	%%%%%%%%%%%%%%%%%%%%%%%%%
= 			\Serif10 			at 8pt
= 		\SerifBold10 		at 8pt
= 	\SerifItalics10 	at 8pt
=	\SerifSlanted10 	at 8pt
=\SerifBoldItalics10	at 8pt
= 		\SansSerif10 		at 8pt
= 	\SansSerifBold10 	at 8pt
=\SansSerifItalics10 at 8pt
=\SansSerifSlanted10 at 8pt
=			\Math10				at 8pt
=		\MathBold10			at 8pt
=			\Symbols10			at 8pt
=		\MoreSymbols10		at 8pt
=		\Typewriter10		at 8pt
=			\Gothic10			at 8pt
=			\Double10			at 8pt

%%%	10		%%%%%%%%%%%%%%%%%%%%%%%%%%%%%%%%%%%%%
%%%	%%%%%%%%%%%%%%%%%%%%%%%%%%%%%
= 			\Serif10 			at 10pt
= 		\SerifBold10 		at 10pt
= 		\SerifItalics10 	at 10pt
=		\SerifSlanted10 	at 10pt
=	\SerifBoldItalics10	at 10pt
= 		\SansSerif10 		at 10pt
= 	\SansSerifBold10 	at 10pt
= 	\SansSerifItalics10 at 10pt
= 	\SansSerifSlanted10 at 10pt
=				\Math10				at 10pt
=			\MathBold10			at 10pt
=			\Symbols10			at 10pt
=		\MoreSymbols10		at 10pt
=		\Typewriter10		at 10pt
=			\Gothic10			at 10pt
=			\Double10			at 10pt
=			\Relazioni10			at 10pt

%%%	12		%%%%%%%%%%%%%%%%%%%%%%%%%%%%%%%%%%%%%
%%%	%%%%%%%%%%%%%%%%%%%%%%%%%%%%%
= 				\Serif10 			at 12pt
= 			\SerifBold10 		at 12pt
= 		\SerifItalics10 	at 12pt
=		\SerifSlanted10 	at 12pt
=	\SerifBoldItalics10	at 12pt
= 			\SansSerif10 		at 12pt
= 		\SansSerifBold10 	at 12pt
= 	\SansSerifItalics10 at 12pt
= 	\SansSerifSlanted10 at 12pt
=				\Math10				at 12pt
=			\MathBold10			at 12pt
=			\Symbols10			at 12pt
=		\MoreSymbols10		at 12pt
=			\Typewriter10		at 12pt
=				\Gothic10			at 12pt
=				\Double10			at 12pt

%%%	14		%%%%%%%%%%%%%%%%%%%%%%%%%%%%%%%%
= 			\Serif10 			at 14pt
= 		\SerifBold10 		at 14pt
= 	\SerifItalics10 	at 14pt
=		\SerifSlanted10 	at 14pt
=	\SerifBoldItalics10	at 14pt
= 		\SansSerif10 		at 14pt
= 	\SansSerifBold10 	at 14pt
= \SansSerifSlanted10 at 14pt
= \SansSerifItalics10 at 14pt
=				\Math10				at 14pt
=			\MathBold10			at 14pt
=			\Symbols10			at 14pt
=		\MoreSymbols10		at 14pt
=		\Typewriter10		at 14pt
=			\Gothic10			at 14pt
=			\Double10			at 14pt

%% Styles	%%%%%%%%%%%%%%%%%%%%%%%%%%%%%%%%%%%%%%%%%%%%%
%% %%%%%%%%%%%%%%%%%%%%%%%%%%%%%%%%%%%%%%%%%%%%%
\def\NormalStyle{\parindent=5pt\parskip=3pt\normalbaselineskip=14pt%
\def\nt{\tenSerif}%
\def\rm{\fam0\tenSerif}%
\textfont0=\tenSerif\scriptfont0=\sevenSerif\scriptscriptfont0=\fiveSerif%text(\tenrm)
\textfont1=\tenMath\scriptfont1=\sevenMath\scriptscriptfont1=\fiveMath%math(\tenmi)
\textfont2=\tenSymbols\scriptfont2=\sevenSymbols\scriptscriptfont2=\fiveSymbols%symbol(\tensy)
\textfont3=\tenMoreSymbols\scriptfont3=\sevenMoreSymbols\scriptscriptfont3=\fiveMoreSymbols%ex(tenex)
\textfont\itfam=\tenSerifItalics\def\it{\fam\itfam\tenSerifItalics}%
\textfont\slfam=\tenSerifSlanted\def\sl{\fam\slfam\tenSerifSlanted}%
\textfont\ttfam=\tenTypewriter\def\tt{\fam\ttfam\tenTypewriter}%
\textfont\bffam=\tenSerifBold%
\def\bf{\fam\bffam\tenSerifBold}\scriptfont\bffam=\sevenSerifBold\scriptscriptfont\bffam=\fiveSerifBold%
\def\cal{\tenSymbols}%
\def\greekbold{\tenMathBold}%
\def\gothic{\tenGothic}%
\def\Bbb{\tenDouble}%
\def\LieFont{\tenSerifItalics}%
\nt\normalbaselines\baselineskip=14pt%
}

%%%%%%%%%%%%%%%%%%%%%%%%%%%%%%%%%%%%%%%%%%%%%%%%%%%%%%%
%%%%%%%%%%%%%%%%%%%%%%
\def\TitleStyle{\parindent=0pt\parskip=0pt\normalbaselineskip=15pt%
\def\nt{\fourteenSansSerifBold}%
\def\rm{\fam0\fourteenSansSerifBold}%
\textfont0=\fourteenSansSerifBold\scriptfont0=\tenSansSerifBold\scriptscriptfont0=\eightSansSerifBold%text(\fourteenrm)
\textfont1=\fourteenMath\scriptfont1=\tenMath\scriptscriptfont1=\eightMath%math(\fourteenmi)
\textfont2=\fourteenSymbols\scriptfont2=\tenSymbols\scriptscriptfont2=\eightSymbols%symbol(\fourteensy)
\textfont3=\fourteenMoreSymbols\scriptfont3=\tenMoreSymbols\scriptscriptfont3=\eightMoreSymbols%ex(fourteenex)
\textfont\itfam=\fourteenSansSerifItalics\def\it{\fam\itfam\fourteenSansSerifItalics}%
\textfont\slfam=\fourteenSansSerifSlanted\def\sl{\fam\slfam\fourteenSerifSansSlanted}%
\textfont\ttfam=\fourteenTypewriter\def\tt{\fam\ttfam\fourteenTypewriter}%
\textfont\bffam=\fourteenSansSerif%
\def\bf{\fam\bffam\fourteenSansSerif}\scriptfont\bffam=\tenSansSerif\scriptscriptfont\bffam=\eightSansSerif%
\def\cal{\fourteenSymbols}%
\def\greekbold{\fourteenMathBold}%
\def\gothic{\fourteenGothic}%
\def\Bbb{\fourteenDouble}%
\def\LieFont{\fourteenSerifItalics}%
\nt\normalbaselines\baselineskip=15pt%
}

%%%%%%%%%%%%%%%%%%%%%%%%%%%%%%%%%%%%%%%%%%%%%%%%%%%%%%%%
%%%%%%%%%%%%%%%%%%%%%
\def\PartStyle{\parindent=0pt\parskip=0pt\normalbaselineskip=15pt%
\def\nt{\fourteenSansSerifBold}%
\def\rm{\fam0\fourteenSansSerifBold}%
\textfont0=\fourteenSansSerifBold\scriptfont0=\tenSansSerifBold\scriptscriptfont0=\eightSansSerifBold%text(\fourteenrm)
\textfont1=\fourteenMath\scriptfont1=\tenMath\scriptscriptfont1=\eightMath%math(\fourteenmi)
\textfont2=\fourteenSymbols\scriptfont2=\tenSymbols\scriptscriptfont2=\eightSymbols%symbol(\fourteensy)
\textfont3=\fourteenMoreSymbols\scriptfont3=\tenMoreSymbols\scriptscriptfont3=\eightMoreSymbols%ex(fourteenex)
\textfont\itfam=\fourteenSansSerifItalics\def\it{\fam\itfam\fourteenSansSerifItalics}%
\textfont\slfam=\fourteenSansSerifSlanted\def\sl{\fam\slfam\fourteenSerifSansSlanted}%
\textfont\ttfam=\fourteenTypewriter\def\tt{\fam\ttfam\fourteenTypewriter}%
\textfont\bffam=\fourteenSansSerif%
\def\bf{\fam\bffam\fourteenSansSerif}\scriptfont\bffam=\tenSansSerif\scriptscriptfont\bffam=\eightSansSerif%
\def\cal{\fourteenSymbols}%
\def\greekbold{\fourteenMathBold}%
\def\gothic{\fourteenGothic}%
\def\Bbb{\fourteenDouble}%
\def\LieFont{\fourteenSerifItalics}%
\nt\normalbaselines\baselineskip=15pt%
}

%%%%%%%%%%%%%%%%%%%%%%%%%%%%%%%%%%%%%%%%%%%%%%%%%%%%%%%%
%%%%%%%%%%%%%%%%%%%%%
\def\ChapterStyle{\parindent=0pt\parskip=0pt\normalbaselineskip=15pt%
\def\nt{\fourteenSansSerifBold}%
\def\rm{\fam0\fourteenSansSerifBold}%
\textfont0=\fourteenSansSerifBold\scriptfont0=\tenSansSerifBold\scriptscriptfont0=\eightSansSerifBold%text(\fourteenrm)
\textfont1=\fourteenMath\scriptfont1=\tenMath\scriptscriptfont1=\eightMath%math(\fourteenmi)
\textfont2=\fourteenSymbols\scriptfont2=\tenSymbols\scriptscriptfont2=\eightSymbols%symbol(\fourteensy)
\textfont3=\fourteenMoreSymbols\scriptfont3=\tenMoreSymbols\scriptscriptfont3=\eightMoreSymbols%ex(fourteenex)
\textfont\itfam=\fourteenSansSerifItalics\def\it{\fam\itfam\fourteenSansSerifItalics}%
\textfont\slfam=\fourteenSansSerifSlanted\def\sl{\fam\slfam\fourteenSerifSansSlanted}%
\textfont\ttfam=\fourteenTypewriter\def\tt{\fam\ttfam\fourteenTypewriter}%
\textfont\bffam=\fourteenSansSerif%
\def\bf{\fam\bffam\fourteenSansSerif}\scriptfont\bffam=\tenSansSerif\scriptscriptfont\bffam=\eightSansSerif%
\def\cal{\fourteenSymbols}%
\def\greekbold{\fourteenMathBold}%
\def\gothic{\fourteenGothic}%
\def\Bbb{\fourteenDouble}%
\def\LieFont{\fourteenSerifItalics}%
\nt\normalbaselines\baselineskip=15pt%
}

%%%%%%%%%%%%%%%%%%%%%%%%%%%%%%%%%%%%%%%%%%%%%%%%%%%%%%%%
%%%%%%%%%%%%%%%%%%%%%
\def\SectionStyle{\parindent=0pt\parskip=0pt\normalbaselineskip=13pt%
\def\nt{\twelveSansSerifBold}%
\def\rm{\fam0\twelveSansSerifBold}%
\textfont0=\twelveSansSerifBold\scriptfont0=\eightSansSerifBold\scriptscriptfont0=\eightSansSerifBold%text(\fourteenrm)
\textfont1=\twelveMath\scriptfont1=\eightMath\scriptscriptfont1=\eightMath%math(\fourteenmi)
\textfont2=\twelveSymbols\scriptfont2=\eightSymbols\scriptscriptfont2=\eightSymbols%symbol(\fourteensy)
\textfont3=\twelveMoreSymbols\scriptfont3=\eightMoreSymbols\scriptscriptfont3=\eightMoreSymbols%ex(fourteenex)
\textfont\itfam=\twelveSansSerifItalics\def\it{\fam\itfam\twelveSansSerifItalics}%
\textfont\slfam=\twelveSansSerifSlanted\def\sl{\fam\slfam\twelveSerifSansSlanted}%
\textfont\ttfam=\twelveTypewriter\def\tt{\fam\ttfam\twelveTypewriter}%
\textfont\bffam=\twelveSansSerif%
\def\bf{\fam\bffam\twelveSansSerif}\scriptfont\bffam=\eightSansSerif\scriptscriptfont\bffam=\eightSansSerif%
\def\cal{\twelveSymbols}%
\def\bg{\twelveMathBold}%
\def\gothic{\twelveGothic}%
\def\Bbb{\twelveDouble}%
\def\LieFont{\twelveSerifItalics}%
\nt\normalbaselines\baselineskip=13pt%
}

%%%%%%%%%%%%%%%%%%%%%%%%%%%%%%%%%%%%%%%%%%%%%%%
\def\SubSectionStyle{\parindent=0pt\parskip=0pt\normalbaselineskip=13pt%
\def\nt{\twelveSansSerifItalics}%
\def\rm{\fam0\twelveSansSerifItalics}%
\textfont0=\twelveSansSerifItalics\scriptfont0=\eightSansSerifItalics\scriptscriptfont0=\eightSansSerifItalics%
\textfont1=\twelveMath\scriptfont1=\eightMath\scriptscriptfont1=\eightMath%
\textfont2=\twelveSymbols\scriptfont2=\eightSymbols\scriptscriptfont2=\eightSymbols%
\textfont3=\twelveMoreSymbols\scriptfont3=\eightMoreSymbols\scriptscriptfont3=\eightMoreSymbols%
\textfont\itfam=\twelveSansSerif\def\it{\fam\itfam\twelveSansSerif}%
\textfont\slfam=\twelveSansSerifSlanted\def\sl{\fam\slfam\twelveSerifSansSlanted}%
\textfont\ttfam=\twelveTypewriter\def\tt{\fam\ttfam\twelveTypewriter}%
\textfont\bffam=\twelveSansSerifBold%
\def\bf{\fam\bffam\twelveSansSerifBold}\scriptfont\bffam=\eightSansSerifBold\scriptscriptfont\bffam=\eightSansSerifBold%
\def\cal{\twelveSymbols}%
\def\greekbold{\twelveMathBold}%
\def\gothic{\twelveGothic}%
\def\Bbb{\twelveDouble}%
\def\LieFont{\twelveSerifItalics}%
\nt\normalbaselines\baselineskip=13pt%
}

%%%%%%%%%%%%%%%%%%%%%%%%%%%%%%%%%%%%%%%%%%%%%%%%%%%%%%%%%%%
%%%%%%%%%%%%%%%%%%
\def\AuthorStyle{\parindent=0pt\parskip=0pt\normalbaselineskip=14pt%
\def\nt{\tenSerif}%
\def\rm{\fam0\tenSerif}%
\textfont0=\tenSerif\scriptfont0=\sevenSerif\scriptscriptfont0=\fiveSerif%text(\tenrm)
\textfont1=\tenMath\scriptfont1=\sevenMath\scriptscriptfont1=\fiveMath%math(\tenmi)
\textfont2=\tenSymbols\scriptfont2=\sevenSymbols\scriptscriptfont2=\fiveSymbols%symbol(\tensy)
\textfont3=\tenMoreSymbols\scriptfont3=\sevenMoreSymbols\scriptscriptfont3=\fiveMoreSymbols%ex(tenex)
\textfont\itfam=\tenSerifItalics\def\it{\fam\itfam\tenSerifItalics}%
\textfont\slfam=\tenSerifSlanted\def\sl{\fam\slfam\tenSerifSlanted}%
\textfont\ttfam=\tenTypewriter\def\tt{\fam\ttfam\tenTypewriter}%
\textfont\bffam=\tenSerifBold%
\def\bf{\fam\bffam\tenSerifBold}\scriptfont\bffam=\sevenSerifBold\scriptscriptfont\bffam=\fiveSerifBold%
\def\cal{\tenSymbols}%
\def\greekbold{\tenMathBold}%
\def\gothic{\tenGothic}%
\def\Bbb{\tenDouble}%
\def\LieFont{\tenSerifItalics}%
\nt\normalbaselines\baselineskip=14pt%
}

%%%%%%%%%%%%%%%%%%%%%%%%%%%%%%%%%%%%%%%%%%%%%%%%%%%%%%%%%%%
%%%%%%%%%%%%%%%%%%

%%%%%%%%%%%%%%%%%%%%%%%%%%%%%%%%%%%%%%%%%%%%%%%%%%%%%%%%%%%
%%%%%%%%%%%%%%%%%%
\def\AbstractStyle{\parindent=0pt\parskip=0pt\normalbaselineskip=12pt%
\def\nt{\eightSerif}%
\def\rm{\fam0\eightSerif}%
\textfont0=\eightSerif\scriptfont0=\sevenSerif\scriptscriptfont0=\fiveSerif%text(\tenrm)
\textfont1=\eightMath\scriptfont1=\sevenMath\scriptscriptfont1=\fiveMath%math(\tenmi)
\textfont2=\eightSymbols\scriptfont2=\sevenSymbols\scriptscriptfont2=\fiveSymbols%symbol(\tensy)
\textfont3=\eightMoreSymbols\scriptfont3=\sevenMoreSymbols\scriptscriptfont3=\fiveMoreSymbols%ex(tenex)
\textfont\itfam=\eightSerifItalics\def\it{\fam\itfam\eightSerifItalics}%
\textfont\slfam=\eightSerifSlanted\def\sl{\fam\slfam\eightSerifSlanted}%
\textfont\ttfam=\eightTypewriter\def\tt{\fam\ttfam\eightTypewriter}%
\textfont\bffam=\eightSerifBold%
\def\bf{\fam\bffam\eightSerifBold}\scriptfont\bffam=\sevenSerifBold\scriptscriptfont\bffam=\fiveSerifBold%
\def\cal{\eightSymbols}%
\def\greekbold{\eightMathBold}%
\def\gothic{\eightGothic}%
\def\Bbb{\eightDouble}%
\def\LieFont{\eightSerifItalics}%
\nt\normalbaselines\baselineskip=12pt%
}

%%%%%%%%%%%%%%%%%%%%%%%%%%%%%%%%%%%%%%%%%%%%%
\def\RefsStyle{\parindent=0pt\parskip=0pt%
\def\nt{\eightSerif}%
\def\rm{\fam0\eightSerif}%
\textfont0=\eightSerif\scriptfont0=\sevenSerif\scriptscriptfont0=\fiveSerif%text(\tenrm)
\textfont1=\eightMath\scriptfont1=\sevenMath\scriptscriptfont1=\fiveMath%math(\tenmi)
\textfont2=\eightSymbols\scriptfont2=\sevenSymbols\scriptscriptfont2=\fiveSymbols%symbol(\tensy)
\textfont3=\eightMoreSymbols\scriptfont3=\sevenMoreSymbols\scriptscriptfont3=\fiveMoreSymbols%ex(tenex)
\textfont\itfam=\eightSerifItalics\def\it{\fam\itfam\eightSerifItalics}%
\textfont\slfam=\eightSerifSlanted\def\sl{\fam\slfam\eightSerifSlanted}%
\textfont\ttfam=\eightTypewriter\def\tt{\fam\ttfam\eightTypewriter}%
\textfont\bffam=\eightSerifBold%
\def\bf{\fam\bffam\eightSerifBold}\scriptfont\bffam=\sevenSerifBold\scriptscriptfont\bffam=\fiveSerifBold%
\def\cal{\eightSymbols}%
\def\greekbold{\eightMathBold}%
\def\gothic{\eightGothic}%
\def\Bbb{\eightDouble}%
\def\LieFont{\eightSerifItalics}%
\nt\normalbaselines\baselineskip=10pt%
}

%%%%%%%%%%%%%%%%%%%%%%%%%%%%%%%%%%%%%%%%%%%%%%%%%%%%%%%%%%%
%%%%%%%%%%%%%%%%%%%

%%%%%%%%%%%%%%%%%%%%%%%%%%%%%%%%%%%%%%%%%%%%%%%%%%%%%%%%%%%%
%%%%%%%%%%%%%%%%%%

%
%    Various Libraries.    Version 1.2.0.beta
%    The best use is to paste all of them into the papers
%     1/8/2005
%

%%%%%%%%%%%%%%%%%%%%%%%%%%%%%%
%%%%%%			Utilities		 %%%%%%
%%%%%%%%%%%%%%%%%%%%%%%%%%%%%%

% Definition modes  %
\def\ModeYes{yes}
\def\ModeNo{no}

\def\ModeUndef{undefined}

%%%%%%%%%%%%

\def\nx{\noexpand}
\def\ni{\noindent}
\def\newpage{\vfill\eject}

\def\ss{\vskip 5pt}
\def\ms{\vskip 10pt}
\def\bs{\vskip 20pt}

 \def\,{\mskip\thinmuskip}
 \def\!{\mskip-\thinmuskip}
 \def\>{\mskip\medmuskip}
 \def\;{\mskip\thickmuskip}

%%%%%%%%%%%%%%%%%%%%%%%%%%%%%%
%%%%%%		Bibliography		 %%%%%%
%%%%%%%%%%%%%%%%%%%%%%%%%%%%%%
%
% Usage:
%	[\SetModeAuto]
% ... 
%	\bib{libro1}{L.Fatibene, ...}
%	\bib{libro2}{L.Fatibene, ...}
% ...
%	(see \ref{libro2} and \ref{libro1})
% ...
% 	\ShowBiblio
%

% Definition modes  %
\def\refsModePost{post}
\def\refsModeAuto{auto}

\def\dbRefsSatusModeOk{ok}
\def\dbRefsSatusModeError{error}
\def\dbRefsSatusModeWarning{warning}

%%%%%%%%%%%%

\newcount\BNUM
\BNUM=0

\def\refs{}

\def\SetModePost{\xdef\refsMode{\refsModePost}}			%	Items are numbered by Citation order
		%	Items are numbered by Insertion order
\SetModePost

\def\dbRefsStatusOk{%
	\xdef\dbRefsStatus{\dbRefsSatusModeOk}%
	\xdef\dbRefsError{\ModeNo}%
	\xdef\dbRefsWarning{\ModeNo}%
	\xdef\dbRefsInfo{\ModeNo}%
}

\def\dbRefs{%
}

\def\dbRefsGet#1{%
	\xdef\found{N}\xdef\ikey{#1}\dbRefsStatusOk%
	\xdef\key{\ModeUndef}\xdef\tag{\ModeUndef}\xdef\tail{\ModeUndef}%
	\dbRefs%
}

\def\NextRefsTag{%
	\global\advance\BNUM by 1%
}
\def\ShowTag#1{{\bf [#1]}}

\def\dbRefsInsert#1#2{%
\dbRefsGet{#1}%
\if\found Y %
   \xdef\dbRefsStatus{\dbRefsSatusModeWarning}%
   \xdef\dbRefsWarning{record is already there}%
   \xdef\dbRefsInfo{record not inserted}%
\else%
   \toks2=\expandafter{\dbRefs}%
   \ifx\refsMode\refsModeAuto \NextRefsTag
    \xdef\dbRefs{%
   	\the\toks2 \nx\xdef\nx\dbx{#1}%
	\nx\ifx\nx\ikey %
		\nx\dbx\nx\xdef\nx\found{Y}%
		\nx\xdef\nx\key{#1}%
		\nx\xdef\nx\tag{\the\BNUM}%
		\nx\xdef\nx\tail{#2}%
	\nx\fi}%
	\global\xdef\refs{\refs \ss\ni[\the\BNUM]\ #2\par}%%%%
   \fi%   	
   \ifx\refsMode\refsModePost 
    \xdef\dbRefs{%
   	\the\toks2 \nx\xdef\nx\dbx{#1}%
	\nx\ifx\nx\ikey %
		\nx\dbx\nx\xdef\nx\found{Y}%
		\nx\xdef\nx\key{#1}%
		\nx\xdef\nx\tag{\ModeUndef}%
		\nx\xdef\nx\tail{#2}%
	\nx\fi}%
   \fi%
\fi%
}

\def\dbRefsEdit#1#2#3{\dbRefsGet{#1}%
\if\found N 
   \xdef\dbRefsStatus{\dbRefsSatusModeError}%
   \xdef\dbRefsError{record is not there}%
   \xdef\dbRefsInfo{record not edited}%
\else%
   \toks2=\expandafter{\dbRefs}%
   \xdef\dbRefs{\the\toks2%
   \nx\xdef\nx\dbx{#1}%
   \nx\ifx\nx\ikey\nx\dbx %
	\nx\xdef\nx\found{Y}%
	\nx\xdef\nx\key{#1}%
	\nx\xdef\nx\tag{#2}%
	\nx\xdef\nx\tail{#3}%
   \nx\fi}%
\fi%
}

\def\bib#1#2{\RefsStyle\dbRefsInsert{#1}{#2}%
	\ifx\dbRefsStatus\dbRefsSatusModeWarning %
		\message{^^J}%
		\message{WARNING: Reference [#1] is doubled.^^J}%
	\fi%
}

\def\ref#1{\dbRefsGet{#1}%
\ifx\found N %
  \message{^^J}%
  \message{ERROR: Reference [#1] unknown.^^J}%
  \ShowTag{??}%
\else%
	\ifx\tag\ModeUndef \NextRefsTag%
		\dbRefsEdit{#1}{\the\BNUM}{\tail}%
		\dbRefsGet{#1}%
		\global\xdef\refs{\refs \ss\ni [\tag]\ \tail\par}%%%%
	\fi
	\ShowTag{\tag}%
\fi%
}

\def\ShowBiblio{\ms\Ensure{\SectionEnsure}%
{\SectionStyle\ni References}%
{\RefsStyle\refs}%
}

%%%%%%%%%%%%%%%%%%%%%%%%%%%%%%
%%%%%%		Label DB			 %%%%%%
%%%%%%%%%%%%%%%%%%%%%%%%%%%%%%
\newcount\CHANGES
\CHANGES=0
\def\AuxFile{7}
\def\PreventDoubleOn{\xdef\PreventDoubleLabel{\ModeYes}}

\PreventDoubleOn

\def\StoreLabel#1#2{\xdef\itag{#2}% Mantiene FileAux e ritorna #2 in \itag
 \ifx\PreModeStatus\ModeNo %
   \message{^^J}%
   \errmessage{You can't use Check without starting with OpenPreMode (and finishing with ClosePreMode)^^J}%
 \else%
   \immediate\write\AuxFile{\nx\dbLabelPreInsert{#1}{\itag}}%     
   \dbLabelGet{#1}%
   \ifx\itag\tag %
   \else%
	\global\advance\CHANGES by 1%
 	\xdef\itag{(?.??)}%
    \fi%
   \fi%
}

\def\PreModeStatus{\ModeNo}

\def\ClosePreMode{\immediate\closeout\AuxFile%
  \ifnum\CHANGES=0%
	\message{^^J}%
	\message{**********************************^^J}%
	\message{**  NO CHANGES TO THE AuxFile  **^^J}%
	\message{**********************************^^J}%
 \else%
	\message{^^J}%
	\message{**************************************************^^J}%
	\message{**  PLAEASE TYPESET IT AGAIN (\the\CHANGES)  **^^J}%
    \errmessage{**************************************************^^ J}%
  \fi%
  \edef\PreModeStatus{\ModeNo}%
}

\def\dbLabelSatusModeOk{ok}

\def\dbLabelSatusModeWarning{warning}

\def\dbLabelStatusOk{%
	\xdef\dbLabelStatus{\dbLabelSatusModeOk}%
	\xdef\dbLabelError{\ModeNo}%
	\xdef\dbLabelWarning{\ModeNo}%
	\xdef\dbLabelInfo{\ModeNo}%
}

\def\dbLabel{%
}

\def\dbLabelGet#1{%
	\xdef\found{N}\xdef\ikey{#1}\dbLabelStatusOk%
	\xdef\key{\ModeUndef}\xdef\tag{\ModeUndef}\xdef\pre{\ModeUndef}%
	\dbLabel%
}

\def\ShowLabel#1{%
 \dbLabelGet{#1}%
 \ifx\tag \ModeUndef %
 	\global\advance\CHANGES by 1%
 	(?.??)%
 \else%
 	\tag%
 \fi%
}

\def\dbLabelPreInsert#1#2{\dbLabelGet{#1}%
\if\found Y %
  \xdef\dbLabelStatus{\dbLabelSatusModeWarning}%
   \xdef\dbLabelWarning{Label is already there}%
   \xdef\dbLabelInfo{Label not inserted}%
   \message{^^J}%
   \errmessage{Double pre definition of label [#1]^^J}%
\else%
   \toks2=\expandafter{\dbLabel}%
    \xdef\dbLabel{%
   	\the\toks2 \nx\xdef\nx\dbx{#1}%
	\nx\ifx\nx\ikey %
		\nx\dbx\nx\xdef\nx\found{Y}%
		\nx\xdef\nx\key{#1}%
		\nx\xdef\nx\tag{#2}%
		\nx\xdef\nx\pre{\ModeYes}%
	\nx\fi}%
\fi%
}

\def\dbLabelInsert#1#2{\dbLabelGet{#1}%
\xdef\itag{#2}%
\dbLabelGet{#1}%
\if\found Y %
	\ifx\tag\itag %
	\else%
	   \ifx\PreventDoubleLabel\ModeYes %
		\message{^^J}%
		\errmessage{Double definition of label [#1]^^J}%
	   \else%
		\message{^^J}%
		\message{Double definition of label [#1]^^J}%
	   \fi%	
	\fi%
   \xdef\dbLabelStatus{\dbLabelSatusModeWarning}%
   \xdef\dbLabelWarning{Label is already there}%
   \xdef\dbLabelInfo{Label not inserted}%
\else%
   \toks2=\expandafter{\dbLabel}%
    \xdef\dbLabel{%
   	\the\toks2 \nx\xdef\nx\dbx{#1}%
	\nx\ifx\nx\ikey %
		\nx\dbx\nx\xdef\nx\found{Y}%
		\nx\xdef\nx\key{#1}%
		\nx\xdef\nx\tag{#2}%
		\nx\xdef\nx\pre{\ModeNo}%
	\nx\fi}%
\fi%
}

%%%%%%%%%%%%%%%%%%%%%%%%%%%%%%
%%%%%%		Numbering			 %%%%%%
%%%%%%%%%%%%%%%%%%%%%%%%%%%%%%

\newcount\PART
\newcount\CHAPTER
\newcount\SECTION
\newcount\SUBSECTION
\newcount\FNUMBER
%\newdimen\TOBOTTOM
%\newdimen\LIMIT

\PART=0
\CHAPTER=0
\SECTION=0
\SUBSECTION=0	
\FNUMBER=0

\def\LastPart{\ModeUndef}
\def\LastChapter{\ModeUndef}
\def\LastSection{\ModeUndef}
\def\LastSubSection{\ModeUndef}
\def\LastClaim{\ModeUndef}
\def\Last{\ModeUndef}

\newdimen\TOBOTTOM
\newdimen\LIMIT

\def\Ensure#1{\ \par\ \immediate\LIMIT=#1\immediate\TOBOTTOM=\the\pagegoal\advance\TOBOTTOM by -\pagetotal%
\ifdim\TOBOTTOM<\LIMIT\newpage \else%
\vskip-\parskip\vskip-\parskip\vskip-\baselineskip\fi}

%%%%%%%%%%%%%%%%%%%%%%%%%%%%%
\def\PartLabel{\the\PART}
\def\NewPart#1{\global\advance\PART by 1%
         \bs\ni{\PartStyle  Part \PartLabel:}
         \bs\ni{\PartStyle #1}\newpage%
         \CHAPTER=0\SECTION=0\SUBSECTION=0\FNUMBER=0%
         \gdef\Left{#1}%
         \global\edef\Last{\PartLabel}%
         \global\edef\LastPart{\PartLabel}%
         \global\edef\LastChapter{\ModeUndef}%
         \global\edef\LastSection{\ModeUndef}%
         \global\edef\LastSubSection{\ModeUndef}%
         \global\edef\LastClaim{\ModeUndef}}
%%%%%%%%%%%%%%%%%%%%%%%%%%%%%
\def\ChapterLabel{\the\CHAPTER}
\def\NewChapter#1{\global\advance\CHAPTER by 1%
         \bs\ni{\ChapterStyle  Chapter \ChapterLabel: #1}\ms%
         \SECTION=0\SUBSECTION=0\FNUMBER=0%
         \gdef\Left{#1}%
         \global\edef\Last{\ChapterLabel}%
         \global\edef\LastChapter{\ChapterLabel}%
         \global\edef\LastSection{\ModeUndef}%
         \global\edef\LastSubSection{\ModeUndef}%
         \global\edef\LastClaim{\ModeUndef}}
%%%%%%%%%%%%%%%%%%%%%%%%%%%%%
\def\SectionEnsure{3cm}
\def\NewSection#1{\Ensure{\SectionEnsure}\gdef\SectionLabel{\the\SECTION}\global\advance\SECTION by 1%
         \ms\ni{\SectionStyle  \SectionLabel.\ #1}\ss%
         \SUBSECTION=0\FNUMBER=0%
         \gdef\Left{#1}%
         \global\edef\Last{\SectionLabel}%
         \global\edef\LastSection{\SectionLabel}%
         \global\edef\LastSubSection{\ModeUndef}%
         \global\edef\LastClaim{\ModeUndef}}
%%%%%%%%%%%%%%%%%%%%%%%%%%%%%
\def\NewAppendix#1#2{\Ensure{\SectionEnsure}\gdef\SectionLabel{#1}\global\advance\SECTION by 1%
         \bs\ni{\SectionStyle  Appendix \SectionLabel.\ #2}\ss%
         \SUBSECTION=0\FNUMBER=0%
         \gdef\Left{#2}%
         \global\edef\Last{\SectionLabel}%
         \global\edef\LastSection{\SectionLabel}%
         \global\edef\LastSubSection{\ModeUndef}%
         \global\edef\LastClaim{\ModeUndef}}
%%%%%%%%%%%%%%%%%%%%%%%%%%%%%
\def\Acknowledgements{\Ensure{\SectionEnsure}\gdef\SectionLabel{}%
         \ms\ni{\SectionStyle  Acknowledgments}\ss%
         \SECTION=0\SUBSECTION=0\FNUMBER=0%
         \gdef\Left{}%
         \global\edef\Last{\ModeUndef}%
         \global\edef\LastSection{\ModeUndef}%
         \global\edef\LastSubSection{\ModeUndef}%
         \global\edef\LastClaim{\ModeUndef}}
%%%%%%%%%%%%%%%%%%%%%%%%%%%%%
\def\SubSectionEnsure{2cm}
\def\SubSectionLabel{\ifnum\SECTION>0 \the\SECTION.\fi\the\SUBSECTION}
\def\NewSubSection#1{\Ensure{\SubSectionEnsure}\global\advance\SUBSECTION by 1%
         \ms\ni{\SubSectionStyle #1}\ss%
         \global\edef\Last{\SubSectionLabel}%
         \global\edef\LastSubSection{\SubSectionLabel}}
%%%%%%%%%%%%%%%%%%%%%%%%%%%%%
\def\SetNumberingModeN{\def\ClaimLabel{(\the\FNUMBER)}}
\def\SetNumberingModeSN{\def\ClaimLabel{(\ifnum\SECTION>0 \SectionLabel.\fi%
      \the\FNUMBER)}}
\def\SetNumberingModeCSN{\def\ClaimLabel{(\ifnum\CHAPTER>0 \the\CHAPTER.\fi%
      \ifnum\SECTION>0 \SectionLabel.\fi%
      \the\FNUMBER)}}

\def\NewClaim{\global\advance\FNUMBER by 1%
    \ClaimLabel%
    \global\edef\LastClaim{\ClaimLabel}%
    \global\edef\Last{\ClaimLabel}}
%%%%%%%%%%%%%%%%%%%%%%%%%%%%%

\def\HideLabels{\xdef\ShowLabelsMode{\ModeNo}}
\HideLabels

\def\fn{\eqno{\NewClaim}} 
\def\fl#1{%
\ifx\ShowLabelsMode\ModeYes%
%\eqno{\relax\hbox to 1cm{\NewClaim\hbox{[#1]}}}%
 \eqno{{\buildrel{\hbox{\AbstractStyle[#1]}}\over{\hfill\NewClaim}}}%
\else%
 \eqno{\NewClaim}%
\fi% 
\dbLabelInsert{#1}{\ClaimLabel}}
\def\fprel#1{\global\advance\FNUMBER by 1\StoreLabel{#1}{\ClaimLabel}%
\ifx\ShowLabelsMode\ModeYes%
%\eqno{\relax\hbox to 1cm{ .\itag\hbox{[#1]}}}%
\eqno{{\buildrel{\hbox{\AbstractStyle[#1]}}\over{\hfill.\itag}}}%
\else%
 \eqno{\itag}%
\fi% 
}

\def\cl#1{\global\advance\FNUMBER by 1\dbLabelInsert{#1}{\ClaimLabel}%
\ifx\ShowLabelsMode\ModeYes%
${\buildrel{\hbox{\AbstractStyle[#1]}}\over{\hfill\ClaimLabel}}$%
\else%
  $\ClaimLabel$%
\fi% 
}
\def\cprel#1{\global\advance\FNUMBER by 1\StoreLabel{#1}{\ClaimLabel}%
\ifx\ShowLabelsMode\ModeYes%
${\buildrel{\hbox{\AbstractStyle[#1]}}\over{\hfill.\itag}}$%
\else%
  $\itag$%
\fi% 
}
%%%%%%%%%%%%%%%%%%%%%%%%%%%%%

\def\Note{\ms\leftskip 3cm\rightskip 1.5cm\AbstractStyle}
\def\endNote{\par\leftskip 2cm\rightskip 0cm\NormalStyle\ss}

%%%%%%%%%%%%%%%%%%%%%%%%%%%%%%
%%%%%%		   Sidebars		          %%%%%%
%%%%%%%%%%%%%%%%%%%%%%%%%%%%%%

\parindent=7pt
\leftskip=2cm
\newcount\SideIndent
\newcount\SideIndentTemp
\SideIndent=0
\newdimen\SectionIndent
\SectionIndent=-8pt

\def\sidebar{\vrule height15pt width.2pt }
\def\endcorner{\hbox{\hbox{\vrule height6pt width.2pt}\vbox to6pt{\vfill\hbox
to4pt{\leaders\hrule height0.2pt\hfill}}}}
\def\begincorner{\hbox{\hbox{\vrule height6pt width.2pt}\vbox to6pt{\hbox
to4pt{\leaders\hrule height0.2pt\hfill}}}}
\def\endbegincorner{\hbox{\vbox to15pt{\endcorner\vskip-6pt\begincorner\vfill}}}
\def\SideShow{\SideIndentTemp=\SideIndent \ifnum \SideIndentTemp>0 
\loop\sidebar\hskip 2pt \advance\SideIndentTemp by-1\ifnum \SideIndentTemp>1 \repeat\fi}

\def\BeginSection{{\vbadness 100000 \par\ni\hskip\SectionIndent%
\SideShow\vbox to 15pt{\vfill\begincorner}}\global\advance\SideIndent by1\vskip-10pt}

\def\EndSection{{\vbadness 100000 \par\ni\global\advance\SideIndent by-1%
\hskip\SectionIndent\SideShow\vbox to15pt{\endcorner\vfill}\vskip-10pt}}

\def\EndBeginSection{{\vbadness 100000\par\ni%
\global\advance\SideIndent by-1\hskip\SectionIndent\SideShow
\vbox to15pt{\vfill\endbegincorner}}%
\global\advance\SideIndent by1\vskip-10pt}

\def\ShowBeginCorners#1{%
\SideIndentTemp =#1 \advance\SideIndentTemp by-1%
\ifnum \SideIndentTemp>0 %
\vskip-15truept\hbox{\kern 2truept\vbox{\hbox{\begincorner}%
\ShowBeginCorners{\SideIndentTemp}\vskip-3truept}}%				
\fi%
}

\def\ShowEndCorners#1{%
\SideIndentTemp =#1 \advance\SideIndentTemp by-1%
\ifnum \SideIndentTemp>0 %
\vskip-15truept\hbox{\kern 2truept\vbox{\hbox{\endcorner}%
\ShowEndCorners{\SideIndentTemp}\vskip 2truept}}%				
\fi%
}

\def\BeginSections#1{{\vbadness 100000 \par\ni\hskip\SectionIndent%
\SideShow\vbox to 15pt{\vfill\ShowBeginCorners{#1}}}\global\advance\SideIndent by#1\vskip-10pt}

\def\EndSections#1{{\vbadness 100000 \par\ni\global\advance\SideIndent by-#1%
\hskip\SectionIndent\SideShow\vbox to15pt{\vskip15pt\ShowEndCorners{#1}\vfill}\vskip-10pt}}

\def\EndBeginSections#1#2{{\vbadness 100000\par\ni%
\global\advance\SideIndent by-#1%
\hbox{\hskip\SectionIndent\SideShow\kern-2pt%
\vbox to15pt{\vskip15pt\ShowEndCorners{#1}\vskip4pt\ShowBeginCorners{#2}}}}%
\global\advance\SideIndent by#2\vskip-10pt}

%%%%%%%%%%%%%%%%%%%%%%%%%%%%%%
%%%%%%		Margin notes		 %%%%%%
%%%%%%%%%%%%%%%%%%%%%%%%%%%%%%

%%%%%%%%%%%%%%%%%%%%%%%%%%%%%

%%%%%%%%%%%%%%%%%%%%%%%%%%%%%

%
%    Macros.    Version 1.2.0.beta
%    The best use is to paste all of them into the papers
%     1/8/2005
%

%%%%%%%%%%%%%%%%%%%%%%%%%%%%%%
%%%%%%			Greek		 %%%%%%
%%%%%%%%%%%%%%%%%%%%%%%%%%%%%%

\def\al{\alpha}
\def\be{\beta}
\def\de{\delta}
\def\ga{\gamma}

\def\ep{\epsilon}

\def\te{\theta}
\def\la{\lambda}
\def\ze{\zeta}
\def\om{\omega}
\def\si{\sigma}
\def\vp{\varphi}

\def\ka{\kappa}

\def\Ga{\Gamma}
\def\Te{\Theta}

\def\Om{\Omega}
\def\Si{\Sigma}

%%%%%%%%%%%%%%%%%%%%%%%%%%%%%%
%%%%%%			Cal			 %%%%%%
%%%%%%%%%%%%%%%%%%%%%%%%%%%%%%
 
 \def\calU{{\hbox{\cal U}}}
 
 \def\calC{{\hbox{\cal C}}}

 \def\calL{{\hbox{\cal L}}}
 \def\smallcalL{{\hbox{{\sevenSymbols  L}}}}
 \def\calE{{\hbox{\cal E}}}

%%%%%%%%%%%%%%%%%%%%%%%%%%%%%%
%%%%%%			gothic		 %%%%%%
%%%%%%%%%%%%%%%%%%%%%%%%%%%%%%

 \def\gotg{{\hbox{\gothic g}}}
 
 		% to prevent \sl redefinition

%%%%%%%%%%%%%%%%%%%%%%%%%%%%%%
%%%%%%			Bbb			 %%%%%%
%%%%%%%%%%%%%%%%%%%%%%%%%%%%%%

 \def\R{{\hbox{\Bbb R}}}

 \def\E{{\hbox{\Bbb E}}}

 \def\F{{\hbox{\Bbb F}}}

 \def\R{{\hbox{\Bbb R}}}

%%%%%%%%%%%%%%%%%%%%%%%%%%%%%%
%%%%%%		MathRoman		 %%%%%%
%%%%%%%%%%%%%%%%%%%%%%%%%%%%%%

\def\Aut{{\hbox{Aut}}}
\def\Div{{\hbox{Div}}}
\def\div{{\hbox{div}}}
\def\ad{{\hbox{ad}}}

\def\SO{{\hbox{SO}}}

\def\GL{{\hbox{GL}}}
\def\det{{\hbox{det}}}

\def\Lor{{\hbox{Lor}}}
\def\Diff{{\hbox{Diff}}}

\def\id{{\hbox{\rm id}}}

\def\supp{{\hbox{supp}}}

%%%%%%%%%%%%%%%%%%%%%%%%%%%%%%
%%%%%%		OtherSymbols		 %%%%%%
%%%%%%%%%%%%%%%%%%%%%%%%%%%%%%
\def\ip{\hbox to4pt{\leaders\hrule height0.3pt\hfill}\vbox to8pt{\leaders\vrule width0.3pt\vfill}\kern 2pt}
% inner product
 
\def\del{\partial}
\def\na{\nabla}

\def\Lie{\hbox{\LieFont \$}}

\def\arr{\rightarrow}

%
%    Format.    Version 1.2.0.beta
%    The best use is to paste all of them into the papers
%     1/8/2005
%

\def\cases#1{\left\{\eqalign{#1}\right.}
%%%%%%%%%%%%%%%%%%%%
\NormalStyle
\SetNumberingModeSN
\PreventDoubleOn

\long\def\title#1{\centerline{\TitleStyle\ni#1}}
\long\def\moretitle#1{\baselineskip18pt\centerline{\TitleStyle\ni#1}}
\long\def\author#1{\ms\centerline{\AuthorStyle by {\it #1}}}

\def\abstract{\ms\leftskip 3cm\rightskip .5cm\AbstractStyle{\bf \ni Abstract:}\ }
\def\endabstract{\par\leftskip 2cm\rightskip 0cm\NormalStyle\ss}

%%%%%%%%%%%%%%%%%%%%%%%%%%%%%
\SetNumberingModeSN
%\ShowLabels

\def\calH{{\hbox{\cal H}}}

\def\frac[#1/#2]{\hbox{$#1\over#2$}}
\def\Frac[#1/#2]{{#1\over#2}}
\def\({\left(}
\def\){\right)}
\def\[{\left[}
\def\]{\right]}
\def\^#1{{}^{#1}_{\>\cdot}}
\def\_#1{{}_{#1}^{\>\cdot}}
\def\Label=#1{{\buildrel {\hbox{\fiveSerif \ShowLabel{#1}}}\over =}}
\def\<{\kern -1pt}

%%%%%%%%  		Collapsable Notes		%%%%%%%%%%%%%%%%%%%%%%%%

\def\ExpandAllCNotes{\long\def\CNote##1{%
\BeginSection%\Margine{{\AbstractStyle To be collapsed}}%
	\Note%
 		##1%
	\endNote% 
\EndSection%
}}
\ExpandAllCNotes
%
% If you want to collapse classes of CNotes independently one of the other, just clone the definition as
%
%	\def\CollapseAllCNotesClassA{\long\def\CNoteClassA##1{}}
%	\def\ExpandAllCNotesClassA{\long\def\CNoteClassA##1{\BeginSection\Note ##1 \endNote\EndSection}}
%	\ExpandAllCNotesClassA
%
%	\def\CollapseAllCNotesClassB{\long\def\CNoteClassB##1{}}
%	\def\ExpandAllCNotesClassB{\long\def\CNoteClassB##1{\BeginSection\Note ##1 \endNote\EndSection}}
%	\ExpandAllCNotesClassB
%
%%%%%%%%%%%%%%%%%%%%%%%%%%%%%%%%%%%%%%%%%%%%%%%%%

%%%%%%%%%%%%			frames 				%%%%%%%%%%%%%%%%%%%

\def\frame#1{\vbox{\hrule\hbox{\vrule\vbox{\kern2pt\hbox{\kern2pt#1\kern2pt}\kern2pt}\vrule}\hrule\kern-4pt}}

\def\Box to #1#2#3{\frame{\vtop{\hbox to #1{\hfill #2 \hfill}\hbox to #1{\hfill #3 \hfill}}}}

%%%%%%%%%%%%%%%%%%%%%%%%%%%%%%%%%%%%%%%%%%%%%%%%%

\bib{EPS}{EPS 
}

\bib{BiMetricTheories}{Komar, BiMetricTheories 
}

\bib{EPS2}{EPS2
}

%\bib{Gravitation}{}
%\bib{Ca}{}

%%%%%%%%%%%%%%%%%%%%%%%%%%%%%%%%%%%%%%%%%%%%%%%%%%%

\def\ubal{\underline{\al}\kern1pt}
\def\obal{\overline{\al}\kern1pt}

\def\ubR{\underline{R}\kern1pt}
\def\obR{\overline{R}\kern1pt}
\def\ubom{\underline{\om}\kern1pt}
\def\obxi{\overline{\xi}\kern1pt}
\def\ubu{\underline{u}\kern1pt}
\def\ube{\underline{e}\kern1pt}
\def\obe{\overline{e}\kern1pt}

%%%%%%%%%%%%%%%%%%%%%%%%%%%%%%%%%%%%%%%%%%%%%%%%%%%
\NormalStyle
%\ShowLabels
%\CollapseAllCNotes

\title{Noether Symmetries and Covariant Conservation Laws}
\moretitle{in Classical, Relativistic and Quantum Physics}

\author{L.~Fatibene,  M.~Francaviglia, S.~Mercadante}

%\address{$^a$ Department of Mathematics, University of Torino (Italy)}
%\moreaddress{$^b$ INFN - Iniziativa Specifica Na12}

\bib{Noether}
    {E.~Noether,
	 {\sl Invariante Variationsprobleme},
	 Nachr.\ d.\ Kšnig.\ Gesellsch.\ d.\ Wiss.\ zu\ Gšttingen,
	 Math-phys.\ Klasse (1918), pp235Ð-257,
	 originally appeared in {\sl Transport Theory and Statistical Physics} {\bf 1}(3) (1971),
	 pp.183-Ð207.
	 An English translation by M.A.~Tavel may be found at arXiv:physics/0503066}

\bib{Libro}
    {L.~Fatibene, M.~Francaviglia,
	 {\sl Natural and Gauge Natural Formalism for Classical Field Theories. A Geometric Perspective
	 Including Spinors and Gauge Theories},
	 Kluver Academic Publisher, Dordrecht (2003) -
	 ISBN:~1-4020-1703-0}

\bib{Steenrod}
    {N.~Steenrod,
	 {\sl Topology of Fibre Bundles},
	 Princeton University Press, Princeton (USA, 1951)}

\bib{KolarPC}{I. Kol\'a\v r, 
	{\sl A geometrical version of the higher order Hamilton formalism in fibred  manifolds}
	J. Geom. Phys. {\bf 1}(2), 1984,  127--137. 
     }
\bib{Garcia}{P.~L.~Garc\'\i a, 
{\sl The Poincar\'e-Cartan Invariant in the Calculusof Variations},
Symposia Math., {\bf 14},	Academic Press, London,   (1976) 219;
P.~L.~Garc\'\i a, J.~Mu\~noz, 
{\sl On the Geometrical Structure of Higher Order Variational Calculus},
in: {\it Proceedings of the
	IUTAM-ISIMM Symposium on Modern Developments in Analytical
	Mechanics}, Torino July 7--11, 1982; S.\ Benenti, M.\
	Francaviglia and A. Lichnerowicz eds., Tecnoprint, Bologna, (1983) 127}

\bib{Ferraris}{M.\ Ferraris, 
{\sl Geometrical Methods in Physics}, 
UJEP Brno, Krupka Ed. 1984, 61; 
     M.\ Ferraris, M. Francaviglia,
     {\sl On the Theory of Formal Connections and Fibered Connections in Fibered Manifolds},
    	Quaderni di Matematica, Universit\`a di Torino,
	Quaderno n.\ {\bf 86}, Torino, 1984}

\bib{Sauders}{D.J.~Saunders, 
{\sl The Geometry of Jet Bundles}, 
Cambridge University Press, Cambridge (UK),	 (1989)}

\bib{Trautman}{A.~Trautman,
{\sl Invariance of Lagrangian Sytems}, in: {\sl Papers in honour of J.L.~Synge},
Clarenden Press (Oxford, UK, 1972), pp.85--99}

\bib{Palais}{R.S.~Palais, C.L.~Terng,
	 {\sl Natural bundles have finite order},
	 Topology {\bf 16} (1977), pp.271--277}

\bib{Gotay}
    {M.J.~Gotay, J.~Isemberg, J.E.~Marsden, R.~Montgomery, J.~\'Sniatycki, P.B.~Yasskin,
     {\sl Momentum Map and Classical Relativistic Fields},
	 (preprint);
	 physics/9801019}

\bib{Augmented}
    {L.~Fatibene, M.~Ferraris, M.~Francaviglia,
	 {\sl Augmented Variational Principles and Relative Conservation Laws in Classical Field Theory},
	 Int.\ J.\ Geom.\ Meth.\ Mod.\ Phys. {\bf 2}(3) (2005), pp.373--392;
	 arXiv:math-ph/0411029}

\bib{Carini}{M.~Carini, L.~Fatibene, M.~Francaviglia,
{\sl Covariant Energy in Mechanics},
Int.\ J.\ Geom.\ Methods Mod.\ Phys., {\bf 4}(6) (2007), pp.907--918}

\bib{ADM}{R.~Arnowitt, S.~Deser, C.~Misner,
{\sl Dynamical Structure and Definition of Energy in General Relativity},
The Physical Review, 116 (1959), pp.1322--1330}

\bib{myADM}{L.Fatibene, M.Ferraris, M.Francaviglia, L.Lusanna,
{\sl ADM Pseudotensors, Conserved Quantities and Covariant Conservation Laws in General Relativity},
(in preparation)}

\bib{Eck}{D.J.~Eck,
	 {\sl Gauge-natural bundles and generalized gauge theories},
     Mem.\ Amer.\ Math.\ Soc.\ {\bf 33}(247) (1981)}

\bib{Kolar}{I.~Kol{\'a}{\v r}, P.W.~Michor, J.~Slov{\'a}k, 
{\sl Natural Operations in Differential Geometry}, 
Springer--Verlag, New York, 1993 USA}

\bib{Holst}{S.~Holst, 
	 {\sl Barbero's Hamiltonian Derived from a Generalized Hilbert-Palatini Action},
     Phys.\ Rev.\ {\bf D53} (1996), p.5966}

\bib{Barbero}
    {F.~Barbero, 
     {\sl Real Ashtekar variables for Lorentzian signature space-time},
     Phys.\ Rev.\ {\bf D51} (1996), p.5507.}

\bib{Immirzi}
    {G.~Immirzi, 
     {\sl Quantum Gravity and Regge Calculus},
     Nucl.\ Phys.\ Proc.\ Suppl.\ {\bf 57} (1997), pp.65--72}

\bib{myBI}
    {L.~Fatibene, M.~Francaviglia, C.~Rovelli, 
     {\sl On a Covariant Formulation of the Barbero-Immirzi Connection}, 
     Class.\ Quant.\ Grav.\ {\bf 24} (2007), pp.3055--3066}

\bib{Rov2}
    {L.~Fatibene, M.~Francaviglia, C.~Rovelli, 
     {\sl Lagrangian Formulation of Ashtekar-Barbero-Immirzi Gravity}
     Class.\ Quant.\ Grav.\ {\bf 24} (2007), pp.4207--4217}

\bib{Jadwisin}
    {L.~Fatibene, M.~Francaviglia, 
     {\sl Deformations of spin structures and gravity},
     Acta Physica Polonica B {\bf 29}(4) (1998), pp.915--928}

\bib{Magnano}{G.~Magnano, M.~Ferraris, M.~Francaviglia,
{\sl Legendre Transformation and Dynamical Structure of Higher-Derivative Gravity},
Class.\ Quant.\ Grav.\ {\bf 7} (1990), pp. 557--570}

\bib{Capozziello}
    {S.~Capozziello, M.~Francaviglia,
     {\sl Extended Theories of Gravitation and Their Cosmological and Astrophysical Applications},
     Gen.\ Rel.\ Grav.\ {\bf 40}(2-3) (2008), pp.357--420}

\bib{Soutiriou}
    {T.P.~Sotiriou, V.~Faraoni,
     {\sl $f(R)$-Theories of Gravity},
	 (2008);
	 arXiv:0805.1726}

\bib{Faraoni}
    {S.~Capozziello, M.~De~Laurentis, V.~Faraoni,
     {\sl A bird's eye view of $f(R)$-gravity}; 
     arXiv:0909.4672}

\bib{DEDM}
    {S.~Capozziello, M.~De~Laurentis, M.~Francaviglia, S.~Mercadante,
	 {\sl From Dark Energy and Dark Matter to Dark Metric},
	 Foundations of Physics, {\bf 39}(10), pp.1161--1176;
	 arXiv:0805.3642v4}

\abstract
We review the Lagrangian formulation of Noether
symmetries (as well as ``generalized Noether symmetries'')
in the framework of Calculus of Variations in Jet Bundles,
with a special attention to so-called ``Natural Theories''
and ``Gauge-Natural Theories,'' that include all relevant
Field Theories and physical applications (from Mechanics
to General Relativity, to Gauge Theories, Supersymmetric
Theories, Spinors and so on).
It is discussed how the use of Poincar\'e-Cartan forms and
decompositions of natural (or gauge-natural) variational
operators give rise to notions such as ``generators of
Noether symmetries,'' energy and reduced energy flow,
Bianchi identities, weak and strong conservation laws,
covariant conservation laws, Hamiltonian-like conservation
laws (such as, e.g., so-called ADM laws in General
Relativity) with emphasis on the physical interpretation
of the quantities calculated in specific cases (energy,
angular momentum, entropy, etc...).
A few substantially new and very recent
applications/examples are presented to better show the power of
the methods introduced: one in Classical Mechanics
(definition of strong conservation laws in a
frame-independent setting and a discussion on the way in
which conserved quantities depend on the choice of an
observer); one in Classical Field Theories (energy and
entropy in General Relativity, in its standard
formulation, in its spin-frame formulation, in its first
order formulation ``{\it \`a la\/} Palatini'' and in its extensions to
Non-Linear Gravity Theories); one in Quantum Field
Theories (applications to conservation laws in Loop
Quantum Gravity via spin connections and Barbero-Immirzi connections).
\endabstract

\NewSection{Introduction}

Symmetries have acquired a central role in Physics.  In Theoretical Physics discrete symmetries
encode most of the intriguing structure of the Standard Model for particles, in Chemistry they
encode for spectroscopic and physical properties of molecules.

Lie groups of transformations encode properties (often enhancing physical interpretation) of
dynamical and Lagrangian systems.  For example, theoretical Relativistic Cosmology is entirely based
on the symmetry ansatz of homogeneity and isotropy (which of course can be later relaxed by using
techniques of Perturbation Theory).  Most of applications of Quantum Gravity (either loopy or
stringy) to Cosmology are entirely based on symmetries since the approach in full generality is
still hindered by massive technical difficulties.

In the beginning of the $20^{th}$ century, Emmy Noether discovered (see \ref{Noether}) a relation
between (continuous families of) symmetries of Lagrangian systems and their first integrals,
i.e.~physical quantities which remain constant during the evolution of the system and are often
related to fundamental physical quantities such as energy, momentum, angular momentum and so on.

Symmetries of mechanical systems (together with their associated conserved quantities) are the basis
of the definition of {\sl integral systems\/} that form a class of systems for which dynamics can be
determined in general by integration.  Generally speaking, knowing a conserved quantity of a
dynamical system allows to reduce the dimension in which the system is defined.  General integration
techniques, such as Hamilton-Jacobi equation, rely in fact entirely on existence of conserved quantities.

Very interesting issues arise when one extends the notion of {\sl transformation\/} to encompass
trasformations depending on velocities (and more generally on accelerations, and so on).  These can
be seen as trasformations on the infinite jet prolongation of the configuration space where the
dynamical system can be seen to be prolonged.  Here infinite dimensionality enters strategically to
require new techniques to extend the procedures which are standard in a finite dimensional arena.

Another interesting setting where symmetries play their role is the framework of Lagrangian field
theories, which are the current basis for any approach to fundamental interations in Physics.  This
corresponds, loosely speaking, to consider a dynamical system with a continuous infinity of degrees
of freedom.  In this context, Noether theorem still holds true, though conservation laws needs to be
interpreted correctly and differently from what is usually done in Mechanics; if the system lives on
a manifold $M$ of dimension $m=\dim(M)$---which usually is identified with the physical spacetime,
or, mathematically speaking, $m$ is the number of independent variables of the equations---Noether
theorem implies the existence of a $(m-1)$-form on $M$ called the {\sl Noether current}.  Such a
current results to be closed along solutions and hence implies a {\sl continuity equation}.  The
{\sl conserved quantities\/} are thence defined as the integrals of such currents on a
$(m-1)$-volume in $M$.  The continuity equation holding for the Noether current relates the changes
of conserved quantities to the flows at the boundary of the region (something entering or escaping
the region) and some residual at singularities.  This setting is particularly suitable for physical
interpretation; it was in fact developed, e.g., to define electric charges in Electromagnetism.
This setting results to go far beyond Electromagnetism and to be the typical situations for (at
least) all fundamental interactions.

In Mechanics $m=1$ (since $M=\R$) and Noether currents are functions; being closed they are constant
along solutions and the situation introduced above is obtained as a special case.  In Field Theory
the role of conservation laws, though certainly important as a support to the physical
interpretation, appears to be weaker than in Mechanics since the mechanical geometrical picture is
lost (we mean, the picture of first integrals each of which determines a level hypersurface of
configuration space so that motion is constrained on the intersection of all these hypersurfaces and
first integrals define a {\sl reduced system} living on it).  Noether currents are $(m-1)$-forms and
define no level hypersurface; on the other hand a Field Theory has infinitely many degrees of
freedom and it is not obvious how to extend mechanical techniques to use conservation laws with the
purpose of reducing the systems accordingly.

On the other hand, the history of Field Theory is strongly entangled with symmetries.  All the
theory of {\sl continuity equations\/} was developed to account in Field Theory of charge
conservation.  When Einstein proposed equations to describe the gravitational field in General
Relativity (GR) their form was obtained on the basis of conservation of energy--momentum tensor
describing matter (as well as, independently, by Hilbert on the basis of his variational principle).

GR itself is a source of fantastic examples of the role of symmetries in Field Theories.  One can
prove general theorems to show that Noether currents are not only closed forms, but even exact
forms along solutions.  This introduces a {\sl superpotential} $\calU$ for each Noether current and
conserved quantities are obtained by surface integrals of the superpotential.  Generalized Stokes
theorem in this context establishes a bridge between conserved quantities and (co)homology of forms
(and hence with topology and global structure of the spacetime manifold) which happens to be central
in the physical interpretation of the model.

Similar situations can be recognized in other areas of Physics such as in Gauge Theories; the same
structures of superpotential forms can be generically recognized in Gauge Theories as well as back
in Mechanics when one wants to use (or needs to use, as it happens for the relativistic material
points) homogeneous formalism (i.e.~treating time and space on equal footing).

Hereafter we shall show some of the relations among symmetries and conservation laws in different areas, preferring a unifying languages that stress similarities.

\NewSection{Geometrical Setting}

We shall here present notation for a geometrical setting for Lagrangian systems which encompasses
both Field Theories and a suitable setting for Mechanics.  We refer the reader to \ref{Libro} for
further and deeper details.

Let $M$ be an (orientable, connected, paracompact) manifold of dimension $m=\dim(M)$ with local
coordinates $x^\mu$ that will be eventually considered as independent variables in the Variational
Calculus.

Let $\calC=(C, M, \pi, F)$ be a bundle over $M$ with projection $\pi$ and standard fiber $F$; see
\ref{Libro}.  Let $y^i$ be local coordinates on $F$ that will eventually represent the dependent
variables of the variational principle representing the dynamics of the Lagrangian system under
consideration.

A vector $w\in TC$ is {\sl vertical} iff $T\pi(w)=0$, i.e.~is expressed in any fibered coordinate
system $(x^\mu, y^i)$ as $w= w^i \del_i$.  The set of all vertical vectors at $p\in C$ is denoted
by $V_p$; the union $V(\pi)=\cup_p V_p\subset TC$ is a sub-bundle of $TC$.

This setting is general enough to encode Field Theories and Mechanics; in particular in Field Theory
$M$ is assumed to represent spacetime (hence usually $m=4$) while in Mechanics one has $M=\R$ (hence
$m=1$) and base coodinates $x^\mu$ usually reduce to $t$.  Configurations of the system are locally
given by assigning the dependent variables as functions of the independent ones.  In Field Theories
this means $y^i(x)$, while in Mechanics $y^i(t)$ locally represent curves in $F$ (which in this case
is called the {\sl configuration space\/} of the system).  Of course global structure of $C$ (or
$F$) encodes the global properties of the system; in particular in Field Theories one considers
trasformation rules of fields with respect to changes of fibrered coordinates; these transformation
rules encode in particular the geometrical character of fields, i.e.~for a Riemannian metric one has
for example
$$
\cases{
&x'^\mu=x'^\mu(x)\cr
&g'_{\mu\nu}= \bar J^\rho_\mu\>g_{\rho\si}\>\bar J^\si_\nu \cr
}
\fn$$
where $ \bar J^\rho_\mu$ denotes the anti-Jacobian of the coordinate change on spacetime, namely of
$x'^\mu=x'^\mu(x)$.

Moreover, these transformation rules encode also global properties, in the technical sense that once
they have been fixed one can prove that there exists a unique bundle $\calC$ (modulo isomorphisms)
having those functions as transition functions.  This is quite satisfactory from the physical
viewpoint since local descriptions given by local observers in terms of local fibered coordinates,
together with transformation rules (which dictate how one can deduce the readings of an observer
knowing the readings of any nearby observer) allow to uniquely describe the physical situation in an
observer-independent and global way.

Global configurations are global sections of the bundle $\calC$, i.e.~maps $\si:M\arr C$ such that
$\pi\circ \si=\id_M$.  They are locally expressed by functions $y^i(x)$ and transformation rules
account for global properties. Usually only local sections exist (i.e.~sections on open subsets 
$U\subset M$) and the existance of global sections of $\calC$ implies topological restrictions 
and/or obstructions.

Once the configuration bundle $\calC$ is given one can define a new bundle, canonically (i.e.~{\sl
functorially\/}) associated to $\calC$, which accounts for derivatives of dependent variables with
respect to independent variables up to some finite order $k$.  This new bundle is called the {\sl
$k$-order prolongation\/} of $\calC$ and it is denoted by $J^k\calC$.  If $(x^\mu, y^i)$ are fibered
coordinates on $\calC$, then $(x^\mu, y^i, y^i_\mu, y^i_{\mu\nu}, \dots)$ are fibered coordinates on
$J^k\calC$, where $y^i_\mu$ stands for the first derivatives, $y^i_{\mu\nu}$ stands for the second
derivatives (and are accordingly assumed to be symmetric in the lower indices $(\mu\nu)$), and so
on.

Any $J^k\calC$ is a bundle over any other $J^{k-s}\calC$ (for any integer $s>0$).  All prolongations
of a bundle define a projective family and one can define its inverse limit, that is denoted by
$J^\infty\calC$.  (This is a bundle in a broader sense, since it is infinite dimensional).

A {Lagrangian of order $k$} is a bundle map $L: J^k\calC \arr A_m(M)$ where $A_m(M)$ is the bundle of $m$-forms over $M$.
Equivalently, the Lagrangian can be seen as a horizontal $m$-form over $J^k\calC$.
The {\sl action\/} is the functional defined as
$$
A_D [\si]= \int_D L\circ j^k \si
\fn$$
for any $m$-region $D\subset M$ and any (local) configuration (i.e., section) $\si$.  {\sl Hamilton
stationary action principle\/} is in this framework a definition for {\sl critical sections},
i.e.~sections that are critical points of the action functional with respect to a canonical class of
deformations.  A {\sl deformation\/} is a vertical vector field $X$ on $\calC$.  Let us denote by
$\Phi_s$ its flow so that we can drag any configuration $\si$ along $X$ defining a $1$-parameter
family of configurations $\si_s= \Phi_s\circ \si$.  The {\sl variation\/} of the action along the
deformation $X$ is defined as
$$
\de_X A_D [\si]= \int_D \Frac[d/ds]\(L\circ j^k \si_s\)|_{s=0}
\fn$$

\Note
{\bf Hamilton stationary action principle}: a configuration $\si$ is critical iff for any compact $m$-region $D\subset M$ and deformation $X$ with 
$\supp(X)\subset D$ one has 
$$
\de_X A_D [\si]=0
\fn$$
\endNote

Equivalently, one can also consider more general deformations only requiring that $X$ vanishes on the boundary $\del D$ toghether with its derivatives up to order $k-1$.
Critical sections can be shown to obey Euler-Lagrangian equations and physically represent allowed configurations (i.e.~configutations which satisfy field equations).
This framework reduces locally to the usual Variational Calculus. 

As we said, in Mechanics one has $M=\R$ and $m=1$.  Being $\R$ a contractible manifolds the bundle
$\calC$ is necessarily trivial, i.e.~diffemorphic to a Cartesian product $\R\times F$; see
\ref{Steenrod}.  Nevertheless, this diffeomorphism is not canonical, but it depends on a {\sl
reference frame\/} which is realized mathematically by a connection on the bundle $\calC$.  A
connection is a family of hyperplanes $H_p\subset T_pC$ such that $H=\cup_{p} H_p$ is a sub--bundle
in $TC$ and at each point $T_pC= V_p\oplus H_p$ (by the way, since the curvature of the connection
is skew and in this case $m=1$, any connection on $\calC$ is {\sl flat}).  A connection in this case
is represented by a distribution $H_p$ of rank $1$ on $C$; being of rank $1$ it is involutive and,
by Fr\"obenious theorem, integrable.  Hence one has defined a foliation in curves (trajectories,
i.e.~unparametrized curves, to be precise) of $C$.  The leaves $\ga_p$ are nowhere vertical and
establish diffemorphisms between any pair of fibers $\pi^{-1}(t_0)$ and $\pi^{-1}(t_1)$; hence they
induce a particular diffeomorphism $t:C\arr \R\times F$, i.e.~a global trivialization, in which
$\ga_p: t\mapsto (t, f_0)$ for some constant $f_0\in F$.  These sections represent {\sl rest\/} for
the global observer associated to the trivialization $t$ defined above.  One has thence two possible
frameworks for Mechanics: one on $\R\times F$, that is a framework for a fixed observer (or
reference frame); and another on $C$, that is potentially independent of the observer.  In both
cases the framework is suitable for describing in particular any holonomic, possibly
time-dependent, Lagrangian system.

Since the Lagrangian is used only to define the action (and then the action itself is used in
Hamilton principle), in all Field Theories one could add to the Lagrangian terms which do not affect
the value of the action functional.  In fact there are forms, called {\sl contact forms}, which
vanish when computed along each configuration of the system; in fact, if a form is contact it
factorizes terms such as $\om^i := d y^i -y^i_\la dx^\la$, $\om^i_\mu := dy^i_\mu -y^i_{\mu\la}
dx^\la$, and so on.

A {\sl Poincar\'e-Cartan form\/} for a Lagrangian system is a form $\Te_L = L + \Om$ on
$J^{2k-1}\calC$ that differs from the Lagrangian $L$ by a contact form $\Om$.  The action can be
written also in terms of the Poincar\'e-Cartan form as
$$
A_D [\si]= \int_D L\circ j^k \si=  \int_D ( j^k \si)^\ast \Te_L 
\fn$$
The contact term $\Om$, which does not affect field equations, solutions and so on, can then be
tuned to enhance the properties related to conservation laws and symmetries.

One requires (besides some other technical requirements; see \ref{Libro}) that for all vertical
fields $X^{(2k-1)}$ of $J^{2k-1}\calC$ the form $i_X d \Te_L$ is contact.  Here $i_X$ denotes the
usual duality between forms and vector fields.  In the next Section we shall see how
Poincar\'e-Cartan forms are adapted to conservation laws and symmetries.  Below we shall see explicit
examples for coordinate expressions of Poincar\'e-Cartan form.

The theory of Poincar\'e-Cartan form was fully developed in the late '70s -- early '80s (see
\ref{KolarPC}, \ref{Garcia}, \ref{Ferraris}).  In Mechanics one can prove that there exists a unique
Poincar\'e-Cartan form at each order.  In Field Theory, there is a unique Poincar\'e-Cartan form in
theories of order $1$, there is a canonical choice for theories of order $2$, while for Field
Theories of order $k\ge 3$ there is a Poincar\'e-Cartan form for each connection on the base
manifold (one needs integration by parts to define it, and one needs {\sl covariant\/} integration
by parts to control globality; different connections define then different Poincar\'e-Cartan forms
when $k\ge 3$).  This non-trivial (and to some extent {\sl unexpected\/}) structure on uniqueness
issue is very beautiful and inspiring, for example for issues related to Hamiltonian formalisms.

\NewSubSection{Field Equations}

Most of Variational Calculus is in fact related (if not completely encoded) in how deformations
defined on the configuration bundle $\calC$ prolong to higher order jet prolongations $J^k\calC$.

A (projectable) vector field $\Xi= \xi^\mu \del_\mu + \xi^i\del_i$ (here {\sl projectable\/} refers
to the fact that the components $\xi^\mu$ are functions of the independent variables only,
i.e.~$\xi^\mu(x)$) can be prolonged to jet bundles to define vector fields $j^k X$ on each
$J^k\calC$.

In general, let $\Phi_s:C\arr C$ be the flow of $X$ which projects onto the flow $\vp_s:M\arr M$ of
the vector field $\xi=\xi^\mu \del_\mu\equiv (\pi)_\ast \Xi $ on $M$.  For any configuration $\si$
one can drag it along the flow by defining $\si_s= \Phi_s\circ \si \circ \vp_s^{-1}$; these can be
easily checked to be configurations as well.  Accordingly, one can define prolonged flows $j^k
\Phi_s: J^k C\arr J^k\calC : j^k_x\si \arr j^k_{\vp_s(x)}\si_s $ together with their infinitesimal
generators $j^k X = \frac[d/ds] j^k \Phi_s |_{s=0}$.

For example for $k=1$ one has that
$$
j^1X= \xi^\mu  \del_\mu + \xi^i\del_i +\(d_\mu \xi^i - d_\mu \xi^\la  y^i_\la\) \del_i^\mu
\fl{jetProlongationVectorFields}$$
is a good (i.e.~global) vector field on $J^1\calC$.  Here $d_\mu = \del_\mu + y^i_\mu \del_i +
y^i_{\mu\la} \del_i^\la + \dots $ denotes the {\sl total derivative\/} operator for (local) functions
on $J^k\calC$.

%For $k=2$ 
%
%\dots {\bf vedi libro} ...

If one restricts to {\sl deformations\/} (i.e., vertical vector fields) prolongations are
$$
\eqalign{
 X&=X^i\del_i \cr
 j^1X&= X^i\del_i  + d_\mu X^i\del_i ^\mu \cr
 j^2X&= X^i\del_i  + d_\mu X^i\del_i ^\mu+ d_{\mu\nu} X^i\del_i ^{\mu\nu} \cr
\dots&\null
 }
\fl{jetProlongationVerticalVectorFields}$$

Field equations of a ($k$-order) Lagrangian $L$ are the differential equations which a
configuration should obey in order to be critical.  The ($k$-order) Lagrangian form is locally
expressed as
$$
L= \calL(x^\la, y^i, y^i_\la, \dots) ds
\fn$$
where $ds$ is the local canonical basis of $m$-forms on $M$ induced by coordinates $x^\mu$ (i.e., 
local {\sl volume\/}).
The deformation is in the form $X= X^i \del_i= \de y^i \del_i$.

By suitable integration by parts one can split the Lagrangian form in two parts, so that
$$
\de_X A_D[\si]= \int_D \E_i X^i ds  + \int_{\del D} \F_i^\mu X^i ds_\mu
\fl{IFVF}$$
where $ds_\mu$ is the local canonical basis of $(m-1)$-forms on $M$ induced by coordinates $x^\mu$.
This splitting is encoded in global bundle morphisms
$$
\cases{
&\E: J^{2k}C \arr V^\ast(C)\otimes A_m(M)\cr
&\F: J^{2k-1}C \arr V^\ast(J^{k-1}C)\otimes A_m(M)\cr
}
\fn$$
where $V^\ast(C)$ are dual to vertical vectors on $C$, while $V^\ast(J^{k-1}C)$ are dual to vertical vectors on $J^{k-1}C$. 
Denoting the dualities by $<\>|\>>$ one has local expressions
$$
\cases{
&<\E| X>= E_i X^i ds\cr
&<\F | j^{k-1}X>= \F_i^\mu X^i ds_\mu\cr
}
\fn$$
For a first order Lagrangian one has for instance
$$
\cases{
&\E_i= \Frac[\del \calL/ \del y^i] - d_\mu \Frac[\del \calL/ \del y^i_\mu]\cr
&\F_i^\mu= \Frac[\del \calL/ \del y^i_\mu]\cr
}
\fn$$

While the {\sl Euler-Lagrange morphism\/} $\E$ is directly related to field equations (and it is
unique at all orders) the {\sl Poincar\'e-Cartan morphism\/} $\F$, that is not unique in general
because of the non-uniqueness of the Poincar\'e-Cartan form, is more properly related to
conservation laws.

Globally one can characterize the morphisms $\E$ and $\F$ by the so-called {\sl first variantion
formula}, that once integrated gives \ShowLabel{IFVF}
$$
<\de L | j^k X>= <\E| X> + \Div <\F | j^{k-1}X>
\fn$$
where $\Div$ is the {\sl divergence operator\/} that corresponds, after evaluation along a
configuration, to exterior differential on forms on $M$.

Let us remark that in Field Theory most Lagrangians are highly degenerate, which means that they do
not depend on all derivatives of fields but just on some (suitable) combinations of them.  Usually,
these combinations are chosen to define some geometrical object with simple transformation laws.  In
this case it is often convenient to proceed by {\sl covariant\/} integration by parts, in order to
control globality of each single term in the equations instead of controlling only the globality of
the whole equation.  Examples will be considered below.

Before turning our attention to symmetries let us mention that the role played by flows above can be
replaced by more general objects.  Let us consider a family of maps $\Phi_s: J^1C\arr C$ that are
locally expressed by
$$
\cases{
&x'^\mu= x'^\mu(x) \cr 
& y'^i= Y^i(x^\la, y^l, y^i_\la)
}
\fn$$
and represent a sort of field transformation depending on the derivatives of field (as it happens in
general for supersymmetries, Backlund tranformations and so on).  Also in this more general
case---in which the trasformation can no longer be interpreted as a geometrical transformation on
the manifold $C$---one can still define an {\sl infinitesimal generator}
$$
\Xi= \xi^\mu(x^\la) \del_\mu + \xi^i(x^\la, y^l, y^i_\la)\del_i
\fl{GenVF}$$
However, this object is no longer a vector field on $C$ (since its components are not local
functions on $C$).  Objects like this are called {\sl generalized vector fields\/} and can be seen
as sections $\Xi$ of the bundle $(\pi^k_0)^\ast(TC)\arr J^kC$ that is defined by pull-back as
follows:
$$
\begindc{\commdiag}[1]
\obj(40,120)[piTC]{$(\pi^k_0)^\ast(TC)$}
\obj(40,80)[JC]{$J^kC$}
\obj(110,120)[TC]{$TC$}
\obj(110,80)[C]{$C$}
\obj(110,40)[M]{$M$}
\mor{C}{M}{$\pi$}
\mor{JC}{C}{$\pi^k_0$}
\mor{TC}{C}{$\tau$}
\mor{piTC}{TC}{}[\leftarrow, \dasharrow]
\mor{piTC}{JC}{}[\leftarrow, \dasharrow]
\cmor((25,80)(10,85)(6,100)(10,117)(15,120))
	\pright(1,100){$\Xi$}
\enddc
$$
By the usual abuse of language the section $\Xi$ has local expression \ShowLabel{GenVF}.

One can then extend the prolongation to generalized vector fields by formally using the same
formulae \ShowLabel{jetProlongationVectorFields} and
\ShowLabel{jetProlongationVerticalVectorFields}.  In the projective limit these families generate
well-defined vector fields on $J^\infty\calC$, while they are, although in a not yet formalized way,
generalized at each finite level $J^k\calC$.

We shall see below examples of such generalized vector fields when they represent symmetries and
hence they define conservation laws.  Let us stress that Emmy Noether was in fact the first to
consider these examples just in her early studies aimed at showing a sort of inverse theorem; each
conservation law can be generated by a suitable generalized symmetry.

\NewSection{Noether Theorem}

Since field equations are mainly encoded in the geometry of jet prolongations (see \ref{Sauders})
Noether theorem can be understood in terms of Lie derivatives (see \ref{Trautman}).

Let $\Xi=\xi^\mu\del_\mu + \xi^i\del_i$ be a (projectable) vector field on the configuration bundle
$\calC$ that projects onto a vector field $\xi=\xi^\mu\del_\mu$ on $M$ and let $\si$ be a
configuration; let us define the {\sl Lie derivative\/} of $\si$ with respect to $\Xi$ to be
$$
\Lie_\Xi \si= (\xi^\mu(x) y^i_\mu -\xi^i(x, y)) \del_i
\fn$$
This is a (generalized) vertical vector field and it accounts for the change of the configuration when dragged along the flow of $\Xi$.
The same expression holds true when $\Xi$ is a generalized vector field by itself. 

This Lie derivative is {\sl natural\/} in the sense that it preserves commutators, i.e.
$$
\Lie_{[\Xi_1, \Xi_2]}= [\Lie_{\Xi_1}, \Lie_{\Xi_2}]
\fl{NaturalityLieBundle}$$
Of course this is true when commutators are considered as the commutators of (possibly generalized)
vector fields on $\calC$.  It is essential to notice here that configurations can be in principle
dragged along the flow of vector fields on $\calC$ only.  In general, there is no dragging in
$\calC$ along vector fields on the base $M$, as one is instead ``phisically'' used to expect in many
physically relevant cases (e.g.~in relativistic applications).  Accordingly, there is no reason to
expect that in general one could define Lie derivative of configurations in $\calC$ along vector
fields on $M$, nor that this can be done in such a way that they are natural, i.e.~they preserve
commutators.

There are however specific bundles in which one could naturally associate a vector field $\hat \xi$
on $\calC$ to each vector field $\xi$ on $M$ (as, e.g., natural bundles or, in a sense,
gauge-natural bundles; see \ref{Libro}.)  For naturality one has that
$$
[\xi, \ze]^{}\hat{}=[\hat \xi, \hat \ze]
\fl{LiftNaturality}$$
One classical example of this situation is on tangent bundles $\calC=TM$; on any tangent bundle one
has in fact a tangent lift of vector fields which preserves commutators.  We have to stress that the
existence of such a natural lift is a property of the bundle $\calC$.  When such a lift is defined
the bundle $\calC$ is in fact called {\sl natural}; one could prove that natural bundles are
associated to (some finite higher order) frame bundle $L^s(M)$ on the base manifold; see
\ref{Palais}.  The tangent bundles are in fact always associated to the standard frame bundle
$L(M)$.

Only on natural bundles one can define dragging along $\xi$ and the corresponding Lie derivative is
defined as
$$
\Lie_{\xi} \si := \Lie_{\hat \xi} \si
\fl{Lift}$$
which is in turn defined as above.  Since the lift $\xi \arr \hat \xi$ is natural one can prove that
this Lie derivative is also natural, i.e.:
$$
\Lie_{[\xi, \ze ]} \si = \Lie_{[\xi,\ze]\hat{}} \si  = [\Lie_{\xi\hat{}},  \Lie_{\ze\hat{}}]\si\equiv  [\Lie_{\xi},  \Lie_{\ze}]\si
\fl{NaturalityLieBasis}$$

We shall present below various examples of this and similar situations.  We wish to stress here that
the naturality \ShowLabel{NaturalityLieBundle} of Lie derivatives on $\calC$ is essential for
applications to conservation laws, while the naturality of the lift \ShowLabel{LiftNaturality}---as
well as the naturality \ShowLabel{NaturalityLieBasis} of the Lie derivatives on $M$---is not.  These
are ``good news,'' since not all Field Theories have a natural lift; examples will be presented in
Gauge Theories and Spinor Theories, in which one could not define a gauge covariant lift that is
natural while conservation laws are still perfectly defined.

Noether theorem asserts a correspondence between Lagrangian symmetries and conservation laws.  A
{\sl Lagrangian symmetry\/} is a (possibly generalized) vector field $\Xi$ on the configuration
bundle $\calC$ such that
$$
<\de L\>|\> j^k \Lie_\Xi \si>=  \Div \( i_\xi L+ <\al | j^r \Lie_\Xi \si>\)
\fl{CovId}$$
for some bundle morphism $\al: J^{k-1} \calC \arr V^{\ast}(J^r \calC)\otimes A_{m-1}(M)$.

By using the first variation formula on the l.h.s. one can easily prove Noether theorem in the form
$$
<\E\>|\>\Lie_\Xi \si> + \Div\(<\F\>|\>j^{k-1}\Lie_\Xi \si>\)=  \Div \( i_\xi L+ <\al | j^r \Lie_\Xi \si>\)
\fn$$
that can be recasted as
$$
\Div\(<\F\>|\>j^{k-1}\Lie_\Xi \si>-  i_\xi L-<\al | j^r \Lie_\Xi \si>\)= - <\E\>|\>\Lie_\Xi \si> 
\fn$$

By direct inspection, the Noether current defined as $\calE=<\F\>|\>j^{k-1}\Lie_\Xi \si>- i_\xi
L-<\al | j^r \Lie_\Xi \si>$ is closed along critical configurations, which manifestly annihilate the
r.h.s. Of course the difficulty in finding conservation laws is here replaced by the difficulties in
finding Lagrangian symmetries $\Xi$.  Examples will also be given below.

Symmetries can be described and treated directly also in terms of the Poincar\'e-Cartan form.
Condition \ShowLabel{CovId} can be written in an equivalent form as follows
$$
\Lie_{j^{2k-1}\Xi} \Te_L= d \al
\fn$$
for some $(m-1)$-form $\al$; i.e., a Lagrangian symmetry leaves the Poincar\'e-Cartan form
invariant.  This characterization of symmetries is beautiful and geometric, and this is a motivation
to define Poincar\'e-Cartan form.

Noether theorem is then simply obtained in the usual form for Lie derivatives of forms, namely
$$
\Lie_{j^{2k-1}\Xi} \Te_L= \Div (i_{j^{2k-1}\Xi} \Te_L)+ i_{j^{2k-1}\Xi} \Div \Te_L
\fn$$
The term $i_{j^{2k-1}\Xi} \Div \Te_L=0$ is nothing but field equations and the Noether current is
given as
$$
\calE= i_{j^{2k-1}\Xi} \Te_L - \al
\fn$$
which, of course, is closed on-shell.

%Leggi di continuit\`a

Conservation laws are thence expressed as (on-shell) closures of Noether currents.  If one chooses
coordinates $x^\mu=(t, x^i)$ adapted to a spacelike foliation of $M$, to mimic what one is used to
do when choosing Cartesian coordinates in Minkowski space, the conservation law is expressed as
$$
\dot \rho + \div j =0
\fn$$
where we set $\rho=\calE^0$ and $j^i=\calE^i$ and $\cdot$ denotes derivative with respect to time
$t$.  These kind of equations are called {\sl continuity equations\/} since $\rho$ is a density and
the quantity
$$
Q=\int_D \rho = \int_D \calE
\fn$$
(for any spacelike $(m-1)$-region $D$ in $M$) is {\sl conserved}, in the sense that its variations
are controlled by the flow of $j$ through the boundary of $D$.  Having said that, conservation laws
$d\calE=0$ are nothing but the covariant form of continuity equations.

%Superpotentials and augmented cohomolies $\arr$ integrali primi.

For a wide class of Field Theories (including Natural and Gauge-Natural Theories) one can also show
that Noether currents are not only closed forms on-shell, but they are also {\sl exact}; see
\ref{Libro}.  One can in fact show, by defining an explicit and algorithmical procedure of covariant
integration by parts, that in those cases the Noether current can be (globally) recasted as
$$
\calE= \tilde \calE + \Div\calU
\fn$$
The $(m-1)$-form  $\tilde \calE$ is called the {\sl reduced current\/} and it vanishes on-shell, while the  
$(m-1)$-form  $\calU$ is called the {\sl superpotential}. Examples will be presented below.
Accordingly, the Noether current is written on-shell as the differential of the superpotential.  

The corresponding conserved quantities are thence defined as surface integrals of the superpotential
$Q= \int_{\Om} \calU$ on $(m-2)$-regions $\Om\subset M$.  This establishes a deep connection between
conserved quantities and cohomology of the spacetime manifold $M$.

The conservation laws involving Noether current (i.e.~$d\calE=0$) hold on on-shell, i.e.~along
critical configurations, and therefore they are called {\sl weak conservation laws}.  When one
defines the superpotential $\calU$ then an equivalent conservation law can be written as
$d(\calE-\tilde\calE)=d(\Div\calU)$; this holds true for each single configuration (also
non-critical).  For such a reason these conservation laws induced by superpotentials are also called
{\sl strong conservation laws}.

In the case of Mechanics one can use the augmented de-Rham sequence
$$
0\arr \R\arr A_0(M)\arr A_1(M) \arr \dots
\fn$$
and the Noether current is an element in $A_0(M)$ that is closed on-shell.  In view of exactness at
the relevant level, Noether currents in Mechanics are therefore {\sl constant\/} on-shell and they
are called {\sl first integrals}.

% Lie algebraic structure remnants in naturality

Finally, let us remark that everything we said above applies also to Lagrangian symmetries $\Xi$;
equivalently, one has a flow of trasformations that leaves the Poincar\'e-Cartan form invariant,
namely the flow of $\Xi$, i.e.~a $1$-parameter group of symmetries.  In many interesting cases, some
of which will be discussed below, Field Theories have a symmetry group of dimension higher than $1$.
 
Noether current, which depends on the symmetry generator, can also be considered as a map from the
Lie algebra of (infinitesimal) symmetries to (on-shell) closed $(m-1)$-forms.  Such a map is usually
called {\sl momentum map} (see \ref{Gotay}) and the naturality with respect to Lie derivation is the
remnant of the group structure of symmetries and their preservation.

Let us stress that symmetries are required to form a group of transformations of configurations, not
necessarily a group of transformations of spacetime.  Sometimes a group of spacetime transformations
naturally induces a group of transformations on configurations---e.g.~when configurations are
represented in terms of spacetime tensors or more generally by geometrical objects---but in general
it does not.  From this viewpoint, Lie derivatives of fields with respect to spacetime vector fields
is an additional (and not necessary) structure; one somehow needs to require naturality
\ShowLabel{NaturalityLieBundle} while the lift \ShowLabel{Lift} and its naturality
\ShowLabel{LiftNaturality} is not always available (nor, in fact, necessary).

\NewSection{Applications to Mechanics}

As examples in Mechanics we shall briefly review some standard examples with the aim of fixing
notation.

The standard framework for first order Mechanics is based on a configuration manifold $Q$ with
points that correspond to system configurations; local coordinates $q^i$ are also known as
Lagrangian coordinates for the system.  The histories of the system are encoded by (parametrized)
curves $\ga:\R\arr Q$ in the configuration space $Q$.

Equivalently, one can define, as discussed above, the configuration bundle $\calC$; 
by means of a reference frame (i.e.~a connection on $\calC$) one can define a global trivialization $\calC\simeq \R\times Q$.
We shall denote fibered coordinates on $\calC$ by $(t, q^i)$.

In this setting histories are in 1-to-1 correspondence with sections of the configuration bundle,
that will be denoted by an abuse of language again by $\ga$.  Since we are restricting to the first
order Mechanics, the Lagrangian should be given on the first jet prolongation $J^1(\R\times Q)$.  By
means of the reference frame we have a special set of motions singled out in the configuration
bundle; these denote ``rest motions'' and can be prolonged (as we saw for any motion) to first jet
bundle.  They define a global isomorphism $J^1(\R\times Q)\simeq \R\times TQ$ since they define what
one has to understand for ``zero velocity'' (of course depending on the reference frame).

Accordingly, we can use fibered coordinates $(t, q^i, u^i)$ on $\R\times TQ$, where $u^i$ denote the Lagrangian velocities.
The dynamics is described by a Lagrangian
$$
L= \calL(t, q^i, u^i) dt
\fn$$
or equivalently by a Poincar\'e-Cartan form
$$
\Te_L=  \calL(t, q^i, u^i) dt + p_i \om^i
\fn$$
where we set $p_i= \Frac[\del \smallcalL/ \del u^i](t, q, u)$ and $\om^i= dq^i - u^i dt$ for the relevant contact $1$-form on $\R\times TQ$.

The Noether theorem presented in general above is generalized in order to encompass the standard one for:

\itemitem{$i$)} a symmetry that is a vertical vector field on the configuration bundle, i.e.~$\Xi=
\xi^i(q) \del_i$ such that its tangent prolomngation $\hat \Xi$ leaves the Lagrangian invariant
($\hat \Xi(L)=0$);

\itemitem{$ii$)} a symmetry that is a (non-vertical) vector field $\Xi$ on the configuration bundle,
e.g.~if the Lagrangian density is independent of time $t$ then $\Xi= \del_t$ is in fact a symmetry;

\itemitem{$iii$)} vector fields that leave {\sl essentially\/} invariant the system (e.g.~when in
the notation introduced above one has $\al\not=0$)

\itemitem{$iv$)} generalized vector fields to obtain first integrals that are not simply linear in
the momenta, as for example the Runge-Lenz vector in Kepler's motion.

\NewSubSection{Case ($i$)}

One can always change fibered coordinates to new fibered coordinates $(t, q^i)$ in which $\Xi=
\del_1$.  That is a symmetry iff the Lagrangian does not depend on $q^1$, i.e.~it is {\sl cyclic}.
In other words these cases correspond to all cases of ignorable coordinates, though $\Xi$ is a
symmetry in all coordinate systems, while $q^1$ is cyclic only when $\Xi= \del_1$.  Thence these
cases are equivalent to a coordinate free notation for cyclic coordinates.

The corresponding first integral is
$$
\calE= \Frac[\del \calL/ \del u^i] \xi^i
\fn$$
that in adapted coordinates corresponds in fact to the momentum $p_1$ conjugated to the cyclic
coordinate $q^1$.

For example, a particle in a constant gravitational field is described by the Lagrangian
$$
L= \frac[m/2]\( \dot x^2+ \dot y^2+ \dot z^2 \) - mg z
\fl{ConstantWeightLag}$$
in Cartesian coordinates $(x, y, z)$. 

The coordinate $x$ ($y$, respectively) is cyclic and the corresponding momentum $p_x= m \dot x$
($p_y= m \dot y$, respectively) is a first integral that corresponds to {\sl linear momentum}.

The vector field $\Xi= x\del_y- y \del_x$ is a symmetry and it corresponds to the fact that in cylindrical coordinates $(r, \te, z)$ the angular coordinate $\te$
is cyclic. The corresponding first integral
$$
\calE= m\(x\dot y- y \dot x\)
\fn$$
corrsponds to the $z$-component of {\sl angular momentum}.

In cylindrical coordinates one has in fact  $\Xi= \del_\te$ and the corresponding first integral is given by
$\calE= mr^2\dot \te$.

\NewSubSection{Case ($ii$)}

When the Lagrangian is independent of time $t$ (e.g.~whenever holonomic constraints are imposed with no explicit time dependence) the vector field $\Xi=\del_t$ is a symmetry and the corresponding first integral is given by
$$
\calH= p_i u^i -L
\fn$$
that corresponds to {\sl mechanical energy}.

In the case of particles in  a constant gravitational field one has
$$
\calH= \frac[m/2]\( \dot x^2+ \dot y^2+ \dot z^2 \) + mg z
\fn$$

\NewSubSection{Case ($iii$)}

For the Lagrangian \ShowLabel{ConstantWeightLag} the vector field $\Xi=\del_z$ fails to leave the Lagrangian invariant;
in fact one has Lie derivatives $\Lie_\Xi x= \Lie_\Xi y=0$ and $\Lie_\Xi z=-1$.
The l.h.s. of the covariance identity \ShowLabel{CovId} reads then as
$$
<\de L\>|\> j^k \Lie_\Xi \si>= - mg \equiv \frac[d/dt]\(-mgt\)
\fn$$
that fails to vanish though it is easily recognized to be the total derivative of a quantity
$\al=-mgt$.  This is for sure physically expected since, in view of gravitational field constancy,
the system is unchanged if everything is translated up along the $z$-axis.  If this is physically
trivial, one should notice that mathematically one needs to consider generalized symmetries even to
encompass these simple examples.

Noether theorem hence applies and the corresponding first integral is
$$
F= m\(- \dot z + gt\)
\fn$$
In this case we know the general solution of the equations of motion ($x= x^1_0 + u^1_0 t$, $y=
x^2_0 + u^2_0 t$, $z= x^3_0 + u^3_0 t - \frac[1/2] gt^2$) and it is easy to show that the quantity
$F\equiv-m u^3_0$ is in fact constant on-shell.

\NewSubSection{Case ($iv$)}

For Kepler system
$$
L=\Frac[1/2]\( \dot r^2+ r^2\dot \te^2 \) +\Frac[ \ka /r]
\fn$$
one can consider the following generalized vector fields
$$
\eqalign{
 \Xi_1&=-r^2\dot\te\sin\te\del_r-(2r\dot \te \cos\te+\dot r\sin\te)\del_\te\cr
 \Xi_2&= r^2\dot\te\cos\te\del_r+(-2r\dot \te\sin\te+\dot r\cos\te)\del_\te}
\fn$$

These are symmetries; in fact, the l.h.s.~of the covariance identity for the first vector reads as
$$
\eqalign{
<\de L\>|\> j^k \Lie_{\Xi_1} \si>=& \Frac[d/dt]\( r^2\sin\te \dot r \dot \te + r^3\dot \te ^2\cos\te +\ka\cos\te\)
}
\fn$$
while for the second vector it reads as
$$
\eqalign{
<\de L\>|\> j^k \Lie_{\Xi_2} \si>=&- \Frac[d/dt]\( r^2\cos\te \dot r \dot \te - r^3\dot \te ^2\sin\te -\ka\cos\te\)
}
\fn$$

The corresponding first integrals $R_A=-\frac[\del \smallcalL/\del u^i]\xi^i_A -\al$ ($A=1,2$) are
$$
\eqalign{
 R_1&= r^3\dot \te^2\cos\te+r^2\dot r\dot\te\sin\te -\ka\cos\te\cr
 R_2&=r^3\dot\te^2\sin\te-r^2\dot r\dot\te\cos\te -\ka\sin\te}
\fn$$
One can easily check that these are the two components of the vector field 
$$
\vec R= R_1\del_x+R_2\del_y= v\times(r\times v)-\ka\del_r
\fn$$
which is called {\sl Laplace vector\/} or {\sl Runge-Lenz vector}.  In other words the vector $\vec
R$ is constant on-shell.  This vector was known in Kepler problems (or, equivalently, in Coulomb
electrostatics) to be related to the fact that perihelia are fixed (and when perturbations are
introduced it relates to the precession of perihelia).

Notice that the vector $\vec R$ is quadratic in the Lagrangian velocities. Since momenta are linear in the Lagrangian velocities
this means that $\vec R$ is quadratic in the momenta. 
Hence, it could not be obtained by ordinary Noether theorem which produces only first integrals linear in the momenta.

%Dipendenza dall'osservatore

\NewSection{GR and Natural Theories}

General Relativity (GR) is based on the principle of general covariance (together with the
equivalence principle).  General covariance is a symmetry requirement; one assumes that spacetime
diffeomorphisms $\Diff(M)$ act on configurations (i.e.~fields are {\sl geometrical objects},
e.g.~tensor fields) and all these transformations induced by spacetime diffeomorphisms are
symmetries for the dynamics.

This assumption combined with Noether theorem has plenty of consequences.  The configuration bundle
is a {\sl natural bundle}, the Lie derivatives are defined with respect to spacetime vector fields
that are all symmetries and they all generate conservation laws.  Moreover, one can show (see
\ref{Libro}) that Noether currents always admit superpotentials and conservation laws are always
defined {\it \`a la\/} Gauss by surface integrals.

Moreover, in such kind of theories the whole set of conservation laws are equivalent to the dynamics of the system.
In fact, one has a $k$-order Lagrangian $L$ which defines field equations $\E=0$ via first variation formula
$$
<\de L| j^k X>= <\E  | X> + \Div<\F | j^{k-1}X>
\fn$$
First variation formula holds in particular for the Lie derivative along each spacetime vector field
$X= L_\xi \si$ and gives conservation laws
$$
\Div\calE= -<\E | \Lie_\xi \si>
\fn$$
for the Noether currents $\calE= <\F | j^{k-1} \Lie_\xi \si> - i_\xi L$.
Then one can define superpotentials $\calU$
$$
\calE=\tilde \calE + \Div \calU
\fn$$
The reduced current is always a combination of field equations.
Thus if one knows Noether currents of a Natural Theory and conserved quantities by means of their superpotentials, 
then one can compute the reduced currents (i.e.~field equations) purely in  terms of conservation laws.

For example, let us consider ``standard GR,'' that is defined as a second order theory on the
configuration bundle $\Lor(M)$ of Lorentian metrics on spacetime $M$, with coordinates $(x^\mu,
g_{\mu\nu})$.  Dynamics is defined by the Hilbert Lagrangian
$$
L_H= \sqrt{g}R ds
\fl{HilbertLag}$$
where $R$ is the scalar curvature of the metric $g$ and  $\sqrt{g}$ denotes the square root of the absolute  
value of the determinant of the metric $g_{\mu\nu}$.

The variation of this Lagrangian along the deformation $X= \de g^{\al\be}\del_{\al\be}$
$$
<\de L_H| j^2 X>= \sqrt{g}\(R_{\al\be}-\frac[1/2] Rg_{\al\be}\)\de g^{\al\be} ds+ \na_\la \( \sqrt{g} g^{\al\be} \de u^\la_{\al\be}\) ds
\fn$$
where $u^\la_{\al\be}= \{g\}^\la_{\al\be}- \de^\la_{(\al} \{g\}^\ep_{\be)\ep}$ and
$\{g\}^\la_{\al\be}$ denotes the Levi-Civita connection of the metric $g$, i.e.~its Christoffel
symbols.

Thence we have
$$
<\E| X>=\sqrt{g}\(R_{\al\be}-\frac[1/2] Rg_{\al\be}\)\de g^{\al\be} ds
\qquad
<\F| j^1 X>= \sqrt{g} g^{\al\be} \de u^\la_{\al\be} ds_\la
\fn$$

% Noether currents
Let us stress that {\sl each\/} spacetime vector field $\xi$---and not only Killing vectors as
sometimes erroneously claimed in the literature---generates conservation laws.  By expanding the Lie
derivatives in terms of the spacetime vectors field one obtains Noether current as
$$
\calE= \sqrt{g}\(\(\frac[3/2]R\^\al{}_\la -R \de^\al_\la\) \xi^\la + \(g^{\be\ga}\de^\al_\la - g^{\al(\ga} \de^{\be)}_\la\) \na_{\be\ga}\xi^\la\) ds_\al
\fn$$
which by suitable covariant integration by parts can be recasted as $\calE=\tilde \calE+ \Div 
\calU$, where we set
%Superpotentials
$$
\tilde \calE= 2\sqrt{g}\( R\^\al{}_\la -\frac[1/2]R \de^\al_\la\) \xi^\la ds_\al
\qquad
\calU= \sqrt{g}\na^{\be} \xi^{\al} ds_{\al\be}
\fn$$
The superpotential $\calU$ is called {\sl Komar potential\/} in honor of Komar who first proposed
it, though originally restricted to a timelike Killing vector $\xi$, while here $\xi$ is instead
{\sl any\/} spacetime vector field.

%Field equations from reduced current
As we discussed above, one can deduce field equations $R_{\al\be}-\frac[1/2] Rg_{\al\be}=0$ from  Noether current $\calE$ and Komar superpotential
$\calU$ by computing reduced current as
$$
\tilde \calE= \calE-\Div \calU
\fn$$
just noticing that this quantity has to vanish for all solutions and all symmetry generators $\xi$.

%Dipendenza dall'osservatore.
This situation, in fact, is completely general for any Natural Theory of any order and any matter
coupling, since it is based on general covariance principle only.  Any Natural Theory comes with a
huge symmetry group, namely $\Diff(M)$, which identifies {\sl intrisically\/} (i.e.~in an observer
independent way) a huge set of conservation laws and conserved quantities.

Problems start when one wants to identify some physically relevant quantity---e.g.~the energy, the
momentum, the angular momentum, \dots---within this intrinsic set of conservation laws.  This is
already a problem in Newtonian Mechanics (see \ref{Augmented}, \ref{Carini}) where it should be
clear from the very beginning that there is no intrinsic notion as {\sl the\/} energy of a system,
in the sense that different observers (even inertial observers) do in fact measure different
energies, momenta, \dots for the same system. This obvious circumstance is often undervalued (or 
even ignored) in current literature on Mechanics, with the consequence of generating a number of 
misunderstandings about conservation low that reverberate and amplify in Field Theories.
This would indeed be a trivial remark if it were not used
sometimes in the literature to argue that conserved quantities in GR must have a non-covariant
genesis.

Such arguments come down to (at least) two different main points:

\itemitem{$i$)} covariant conservation laws are not conservation laws;

\itemitem{$ii$)} covariant conserved quantities would not depend on the observer as physically expected and as pseudotensorial prescriptions do.

Both these points have a long history and can be somehow traced back to Einstein himself.  However,
after almost one century of investigations they can be today shown to be flat wrong at least in some
sense.

\NewSubSection{Covariant conservation laws}

Item $(i)$ comes historically from the observation that covariant conservation laws
$$
\na_\mu \calE^\mu =0
\fn$$
would not reduce to continuity equations (which is what one usually means for ``conservation'') due
to terms depending on the connection $\{g\}^\la_{\al\be}$ used to define the covariant derivatives.
This would be certainly the case if $\calE^\mu$ were components of a vector field.  Then {\sl
covariant conservation laws\/} would be {\sl conservation laws\/} {\it tout court\/} only for those
observers (here identified with coordinate systems) in which such terms vanish (i.e.~when
$\{g\}^\la_{\al\be}=0$).  Only in these cases covariant conservation laws are genuine conservation
laws and they are intrinsically non-covariant since they break down general covariance.

This argument is true for all currents $\calE$ except in one single case: the case in which
$\calE^\mu$ is a vector density of weight $1$. In this case there are two terms depending on the
connection and they cancel out.  In other words, when $\calE^\mu$ is a vector density of weight $1$
the covariant divergence {\sl is\/} automatically identical to the {\sl ordinary divergence}, i.e.
$$
\na_\mu \calE^\mu \equiv d_\mu \calE^\mu
\fn$$
 and they always define true continuity equations for any observer.

Of course, Noether theorem in the form presented here dictates for Noether currents to be 
$(m-1)$-forms, i.e.~their components $\calE^\mu$ {\sl are\/} in fact 
vector densities of weight $1$.
Hence it always produces authentic conservation laws which are at the same time {\sl covariant}.

\NewSubSection{Observers}

Item $(ii)$ comes from the belief that covariant conservation laws would necessarily define
conserved quantities independent of the observer.
This is certainly true for the whole set of conservation quantities; 
it is defined covariantly and all observers agree on it.

However, each observer can then be asked to identify within the set of all conserved quantities
which one represents a physically relevant quantity, e.g.~the energy.  It is not the notion of first
integral to be observer dependent; a quantity either changes or not during the evolution of the
system and all observers agree on it.  But it is rather the {\sl energy\/} that is related (in
Newtonian Mechanics, but also in Special Relativity) to time translations; since in SR {\sl time\/}
depends on the observer, the notion of {\sl energy\/} depends on the observer, accordingly.

In GR each observer (which is identified with a local coordinate system together with a protocol to
synchronize clocks at different points, that foliates spacetime with leaves representing synchronous
event sets; such foliations are known as ADM-foliations, see \ref{ADM}) comes with a timelike vector
field $\xi$.  The Noether quantity associated to such a timelike vector field (no request about it
being Killing; see \ref{myADM}) is defined to be the {\sl energy for the observer for which timeflow
is given by $\xi$}.

Since the prescription is covariant, any other observer agrees that this energy is the Noether
quantity associated to that timeflow $\xi$; however, different observers might disagree on $\xi$
being the timeflow and consequently that such a conserved quantity has to be considered to be {\sl
the energy\/} of the system.  On the contrary, they use {\sl their\/} own timeflow $\xi'$ and define
{\sl their own energy}.  In this way, which is the only reasonable way to extend to GR what one is
used to do in SR, despite the prescription is covariant still the {\sl energy\/} is not and still 
it depends on the observer.
 
%Serve sapere cos' l'energia?
It is therefore an open issue whether we really need to know what is the {\sl energy\/} or we could
have a fundamentally good description of the physical world by just using the set of conserved
quantities.  In other words, if energy is just something we are used from Newtonian Physics, or 
rather one
can produce a fundamental description of the physical world without singling out special conserved
quantites to be given special meanings.

% Cosa serve avere prescrizione covariante?
One could also ask whether it is useful to have covariant prescriptions for conserved quantities,
since sooner or later covariance have to be broken in favour of observer-dependent quantities.  Why
one should not be satisfied with coordinate-depending prescriptions such as the ones based on
pseudotensors?  Let us counter--argue that on manifolds there is nothing defined as coordinate
dependent integrals to define conserved quantities from pseudotensors.  Pseudotensors are coordinate
dependent quantities, often their genesis from geometric objects is not clear, they are known in
some coordinate system but they are often unspecified in other coordinates.  Moreover---and for
obvious reasons---they lack of any coherent and (mathematically and physically) intrinsic meaning,
if any may even exist!

We do not claim that pseudotensors should be completely forbidden in Physics, snce they reveal to be
often ``useful;'' but we believe that for a reasonable, global and covariant interpretation their
genesis from some geometric object must be always made explicit and, in any case, they have to be
treated {\it cum grano salis}.  Pseudotensors should be defined starting from geometric objects by
fixing coordinates in order to neglect some term.  This scenario has two good features to be
noticed: it implicitly defines the coordinate systems in which pseudotensors can be used and it
defines its integrals by means of the integral of the original geometric object.  Let us refer to
\ref{myADM} for an example of this coherent strategy.

\NewSubSection{Extended Theories of Gravitation}

As we said above, the structure of GR is not peculiar of the Hilbert Lagrangian
\ShowLabel{HilbertLag}; most of this structure is determined by the symmetry group, i.e.~spacetime
diffeomorphisms, and it is a feature of any generally covariant theory.

Let us here briefly consider the class of {\it Extended Theories of Gravitation} (ETG); see
\ref{Magnano}, \ref{Capozziello}, \ref{Soutiriou}.  This class of theories is used in Cosmology and
Astrophysics in order to model phenomena and observations that are usually related to dark matter
and dark energy; \ref{Faraoni}.

Let us first consider a Lorentzian metric $g_{\mu\nu}$ and a connection $\Ga^\la_{\mu\nu}$ as fundamental fields. The configuration bundle is 
the product $\Lor(M)\times C(M)$ of the bundle of Lorentzian metrics and the bundle of connections on $M$ (here assumed for simplicity with no torsion).
Here connections are in principle independent of the metric field.

Let us denote by $R^\al{}_{\be\mu\nu}$ ($R_{\be\nu}$) the Riemann (Ricci) tensor of the connection
$\Ga$ and by $R=g^{\be\nu} R_{\be\nu}$ the scalar curvature which depends on both the metric and the
connection.  These {\sl curvatures\/} are not to be confused with the curvatures
${}^{(g)}\<\<R^\al{}_{\be\mu\nu}$, ${}^{(g)}\<\<R_{\be\nu}$, ${}^{(g)}\<\<R$ of the metric $g$
alone.  This double affine structure (the one induced by $\Ga$ and the one induced by $g$) opens a
number of issues about physical interpretation; see \ref{DEDM}.

In the notation here introduced let us consider  the Lagrangian
$$
L_f= \sqrt{g} f(R) \> ds
\fl{ETGLag}$$
where $f$ is a {\sl generic\/} analytic function of the scalar curvature.  For the specific case
$f(R)=R$ this reduces to standard Hilbert-Einstein Gravity \ShowLabel{HilbertLag}.

When varying, the metric and connection are deformed independently and there are field equations
from variation of both fields:
$$
\cases{
& f'(R) R_{(\mu\nu)}  -\frac[1/2] f g_{\mu\nu}   = 0\cr
&\na_\la \(\sqrt{g}  f'  g^{\mu\nu}\) =0 \cr
}
\fn$$

The second equation is solved defining a new metric field $h_{\mu\nu}= f'(R) g_{\mu\nu}$ conformal to the original one; the connection is then forced to be the 
Levi-Civita connection of the metric $h$, i.e.~$\Ga^\al_{\be\mu}= \{h\}^\al_{\be\mu}$. 

The trace (by $g^{\mu\nu}$) of the first equation $f'(R) R - 2 R=0$ is called the {\sl master 
equation\/} and (once $f$ is considered a fixed function) 
it is {\sl algebraic\/} in the scalar curvature $R$. In general the master equation allows to solve the scalar curvature as a function of the matter sources; 
here we are considering the vacuum case and $R$ is constant (on $M$) and determined by the zeroes of the master equation (which are finite and generically simple).
When this information is plugged into the first field equation one can show that the metric $h$ 
obeys {\sl modified\/} Einstein equations; the function $f$ can be chosen so that the modified 
dynamics accounts for dark matter/dark energy phenomenology with no need to introduce exotic matter sources, which currently cannot be observed directly. 

Again these models are generally covariant and any spacetime diffeomorphism is a symmetry of the theory; they are Natural Theories.
Any spacetime vector field $\xi$ is a symmetry generator and its Noether current is given by
$$
\eqalign{
\calE=&\sqrt g \(f'(R)\,g^{\al\be}\,\Lie_\xi u^\la_{\al\be} - \xi^\la f(R)\)ds_\la=\cr
=&2\sqrt{h}\( {}^{(h)}\<\<R^\la{}_\ep -\frac[1/4] {}^{(h)}\<\<R \de^\la_\ep\)\xi^\ep ds_\la +\na_\ep \(\sqrt{h} \na^{[\ep}\xi^{\la]}\) ds_\la
}
\fn$$
where we used the master equation, $\Ga=\{h\}$ and the curvature tensors of the conformal metric $h$ are denoted by ${}^{(h)}\<\<R^\al{}_{\be\mu\nu}$, 
${}^{(h)}\<\<R_{\be\nu}$ and ${}^{(h)}\<\<R$.

This correponds to define the reduced current and the superpotential as:
$$
\tilde\calE= 2\sqrt{h}\( {}^{(h)}\<\<R^\la{}_\ep -\frac[1/4] {}^{(h)}\<\<R \de^\la_\ep\)\xi^\ep ds_\la 
\qquad\qquad
\calU= \sqrt{h} \na^{[\ep}\xi^{\la]} ds_{\la\ep}
\fn$$
Notice once again that the reduced current $\tilde\calE$ identically vanishes on-shell and it is equivalent to field equations. 

ETG are also considered in the purely metric formalism, in which one considers the connection $\Ga$
to be $\Ga=\{g\}$ from the beginning, i.e.~at the kinematical level.  The configuration bundle is
simply $\Lor(M)$ and the Lagrangian \ShowLabel{ETGLag} (now considered as a function of the metric
$g$ alone) is second order.  This leads to forth order equations with modified effects with respect
to standard GR.

From this point of view standard GR is shown to be degenerate (it has second order field equations)
with respect to the extensions considered here (which generically have forth order field equations).
Let us remark how the structure of symmetry group and conservation laws is instead similar in
standard GR and ETG. It does not depend in fact on the particular dynamics considered but just on
the principle of general covariance.

\ 

\NewSection{Gauge Theories}

Natural Theories are defined starting from a spacetime manifold $M$ together with its
diffeomorphisms that encode symmetries.  It is well-known that this is suitable for some physical
systems (specifically, Relativistic Theories) while there are more general physical systems
(e.g.~Gauge Theories) which need bigger symmetry groups.  Gauge Theories are needed today to
describe some fundamental interactions and togheter with GR they are enough for Fundamental Physics.

A Gauge Theory depends on a principal bundle $P$ with a structure Lie group $G$ over spacetime $M$.
Principal automorphisms $\Aut(P)$ form a transformation group; this group projects onto $\Diff(M)$,
but $\Diff(M)$ is not in general realised as a subgroup of $\Aut(P)$.  Automorphisms $\Phi\in
\Aut(P)$ are realised locally as changes of coordinates together with an element of $G$ acting at
all spacetime points (which in Physics is called a {\sl local action\/} of $G$).

A Gauge-Natural Theory is a Field Theory in which $\Aut(P)$ acts on configurations generating
Noether symmetries of the dynamics.  This request forces the configuration bundle to be a
gauge-natural bundle (see \ref{Libro}, \ref{Kolar}, \ref{Eck}); all fundamental theories  of
Physics can be cast in this form.

As in Natural Theories, also in Gauge-Natural Theories one can associate a Noether current to any
right--invariant vector field on $P$ (i.e.~to each generator of principal automorphisms).  Noether
currents always allow superpotentials (see \ref{Libro}) and field equations can be obtained purely
from conservation laws.  Yang-Mills theories will be discussed below

\NewSubSection{Yang-Mills theory}

Let us briefly present Yang-Mills theories as a paradigm for Gauge(-Natural) Theories.  Also Dirac
spinors share most of the characteristics of Gauge Theories and can in fact be formulated as a
Gauge-Natural Theory; see \ref{Libro}.

Let us consider a semisimple Lie group $G$ and denote by $\de$ the Cartan-Killing ($\ad$-invariant)
metric on its Lie algebra $\gotg$.  Let $P$ be a principal bundle with structure group $G$ and
$C(P)$ be the bundle of principal connections on $P$.  Automorphisms act on connections so that
$C(P)$ is a gauge-natural bundle.  Once one fixes a $\de$-orthonormal basis $T_A$ in the Lie algebra
$\gotg$ a connection is expressed by coefficients $\om^A_\mu(x)$; the curvature is defined by
$F^A_{\mu\nu}= d_\mu \om^A_\nu - d_\nu \om^A_\mu + c^A{}_{BC}\om^B_\mu\om^C_\nu$, where $c^A{}_{BC}$
are ``structure constants'' of the group $G$.  An automorphism on $P$ is locally expressed as
$$
\cases{
&x'^\mu= x'^\mu(x)\cr
&g'= \vp(x)\cdot g\cr
}
\fn$$  
for some local map $\vp:M\arr G$, $\cdot$ being the product in the group; it acts on the curvature
as
$$
F'^A_{\mu\nu}= \ad^A_B(\vp) F^B_{\rho\si} \bar J_\mu^\rho\bar J_\nu^\si
\fn$$  
where $ \bar J_\mu^\rho$ is the anti-Jacobian of the coordinate change and $\ad^A_B$ is the adjoint action of the group $G$ on its Lie algebra $\gotg$. 

The Yang-Mills Lagrangian
$$
L_{YM}= -\frac[1/4]\sqrt{g} \de_{AB} F^A_{\mu\nu} F^B_{\rho\si} g^{\mu\rho} g^{\nu\si} ds
\fn$$
defines the dynamics and gauge transformations (i.e.~automorphisms in $P$) are symmetries.

%Field equations
The first variation formula of the Lagrangian with respect to the deformation $X= \de g^{\al\be} \del_{\al\be} + \de \om^A_\mu \del_A^\mu$ defines 
the morphisms
$$
<\E| X>= \(\na_\mu \(\sqrt{g} F_A^{\mu\nu}  \) \de \om^B_\nu  + \sqrt{g} H_{\mu\nu} \de g^{\mu\nu} \)ds
\qquad
<\F|X>= -\sqrt{g} F_A^{\mu\nu}   \de \om^B_\nu ds_\mu
\fn$$
where we set $F_A^{\mu\nu}:= \de_{AB} F^B_{\rho\si} g^{\mu\rho} g^{\nu\si}$ and $H_{\mu\nu}:=
F^A_{\mu\la} F_A{}_\nu{}^\la -\frac[1/4] F^A_{\la\si} F_A^{\la\si}g_{\mu\nu}$.  The tensor
$H_{\mu\nu}$ is also known as the energy-momentum tensor of the Yang-Mills field.  If the metric is
freezed, then field equations are $\na_\mu \(\sqrt{g} F_A^{\mu\nu} \)=0$, i.e.~they are the
Yang-Mills equations.  In general, the Yang-Mills Lagrangian is coupled to gravity and $H_{\mu\nu}$
acts as a source for the gravitational field.

%Noether currents
Given a pointwise basis $\rho_A$ of vertical right-invariant vector fields on $P$, symmetry generators are in the form
$\Xi= \xi^\mu(x)\del_\mu + \xi^A(x)\rho_A$. The corresponding Noether current is
$$
\calE=-\sqrt{g} \(H\^\nu{}_\mu \xi^\mu + F_A^{\nu\mu} \na_\mu \xi^A_{V}\) ds_\nu
\fn$$
where we set $\xi^A_{V}:= \xi^A + \om^A_\mu \xi^\mu$ for the vertical part of $\Xi$ with respect to the dynamical connection $\om$.

%superpotentials 
The superpotential is given by
$$
\calU=-\frac[1/2]\sqrt{g} F_A^{\nu\mu} \xi^A_V ds_{\mu\nu}
\fn$$

In the special case $G=U(1)$ (i.e., Maxwell's Electromagnetism) the group is commutative and the
adjoint representation is trivial; the curvature is then a $2$-form on spacetime and it is invariant
with respect to (proper) gauge transformations (i.e.~vertical automorphisms).  In this case
Yang-Mills theory reduces to standard Electromagnetism and the connection $\om_\mu$ represents the
quadripotential of electromagnetic field.  Field equations reduce to Maxwell equations and gauge
charges reduce to electric charges.

\NewSubSection{Hole argument}

There is another important crosspoint among GR, Gauge Theories and their symmetries.  It is called
the {\sl hole argument\/} and it was first discovered in GR by Einstein, though it persists in Gauge
Theories as well.  It is originated when among symmetries of the system there are compact supported
symmetries, i.e.~symmetries that vanish (or became the identity) out of a compact sub-domain of
spacetime $M$.

In GR one can easily define compact supported spacetime diffeomorphisms (and vector fields) on any
paracompact spacetime manifold $M$.  In Gauge Theories one can define compact supported
automorphisms.  Of course in Mechanics it is difficult to find examples of compact supported
symmetries (e.g.~there are no compact supported rotations in space).  The only well-known exception
is curve reparametrization in SR; there Physics is totally represented in terms of trajectories in
spacetime (the so-called {\sl worldlines\/}) and their parametrizations are introduced for technical
reasons only, since they are not endowed with any physical meaning.  Equations of motion should
however induce equations on trajectories; hence they must be invariant with respect to general
reparametrizations, which are thence required to be symmetries of the system.  In this case, it is
easy to produce compact support reparametrizations since the parameter lives in $\R$ which is
paracompact.

The hole argument is based on the fact that symmetries transform solutions into solutions (since
they leave the action invariant).  One can show in an elementary way that compact supported
symmetries contradict directly and essentially Cauchy theorem (i.e.~determinism).  In fact, given a
solution $\si$ and acting on it by a compact supported symmetry $\Phi$ one obtains a new solution
$\si'=\Phi_\ast \si$.  The two solutions $\si$ and $\si'$ are identical everywhere except within the
compact support $D$ of $\Phi$.  One has thence two different solutions which are exactly the same
far in the past (in particular on a spacelike hypersurface where initial conditions can be imposed
and which does not intersect $D\subset M$) while they differ at some point (in $D$).

This is a very general property of possible dynamics with symmetry groups that contain compact
supported symmetries.  Since in Physics one does not want to give up determinism too easily the only
other possible way out is to assume that configurations which differ by compact supported symmetries
(e.g.~$\si$ and $\si'$ above) represent in fact the same physical situation.  One physical situation
can be represented by many different mathematical representations, i.e.~physical situations are not
in 1-to-1 correspondence with configurations but with equivalence classes of configurations.  In
other words, physical situations are realized by some suitable quotient of configuration space.

Dynamics must therefore respect the quotient (that is equivalent to assume that a change of
representative in equivalence class is a dynamical symmetry).  On the other way, just because of the
structure of the symmetry group, one cannot hope to observe any difference among different
representatives of the same class of equivalence.  In other words, symmetries not only constrain
physically reasonable dynamics but they also impose strong constraints on what can be physically
observed.  Accordingly, in all cases that are relevant to fundamental interaction Physics, the
dynamics is degenerate.

This is, to the best of our knowledge, the widest definition of what one could mean by {\sl Gauge
Theories}.  A Gauge Theory could be defined as a Field Theory with a symmetry group that contains
compact supported symmetries; accordingly, its dynamics is degenerate and it induces a deterministic
dynamics on the space of gauge classes of configurtations that are identified with physical
situations.

For this reason physical gravitational fields in GR are not described in terms of ``metrics'' but 
rather in
terms of ``geometries'' (i.e.~equivalence classes of metrics with respect to spacetime
diffeomorphisms).  In Gauge Theories Yang-Mills fields are identified with gauge classes of
connections on $P$.  In SR motions are unparametrized worldlines.

\NewSection{Frame-Affine Formalism for GR}

In order to provide an example of concrete application of our formalism here introduced in action we
shall here consider the application to the so called {\sl Holst's action principle\/} (see
\ref{Holst}) which is used as an equivalent formulation of GR suitable for developing LQG through
the use of the Barbero-Immirzi connection (see \ref{Barbero}, \ref{Immirzi}, \ref{myBI}, \ref{Rov2}
as well as references quoted therein).

Let $M$ be a $m$--dimensional manifold (which will be required to allow global metrics of signature $\eta=(r,s)$, with $m=r+s$).
Let us denote by $x^\mu$ local coordinates on $M$, which induce a basis $\del_\mu$ of tangent spaces;
let $L(M)$ denote the {\sl general frame bundle\/} of $M$ and set $(x^\mu, V_a^\mu)$ for fibered coordinates on $L(M)$.
We can define a right-invariant basis for vertical vectors on $L(M)$
$$
\rho^\mu_\nu= V^\mu_a \Frac[\del/\del V^\nu_a]
\fn$$
The general frame bundle is natural (see \ref{Kolar}), hence any spacetime vector field $\xi=\xi^\mu\del_\mu$ defines a natural lift on $L(M)$
$$
\hat\xi=\xi^\mu\>\del_\mu+ \del_\mu\xi^\nu \>\rho^\mu_\nu
\fn$$
We stress that the lift vector field $\hat \xi$ is global whenever $\xi$ is global.

A connection on $L(M)$ is denoted  by $\Ga^\al_{\be\mu}$ and it defines a lift
$$
\Ga:TM\arr TL(M):\xi^\mu\del_\mu\mapsto  \xi^\mu \( \del_\mu - \Ga^\al_{\be\mu} \rho^\be_\al\)
\fn$$
This lift does not in general preserve commutators, unless the connection is flat.

%If a metric $g=g_{\mu\nu} \>dx^\mu\otimes dx^\nu$ is given on $M$ then its Christoffel symbols define the Levi-Civita connection of the metric.
%Such a connection is torsionless (i.e.~symmetric in lower indices) and {\it compatible with the metric}, i.e.~such that $\na_\mu g_{\al\be}=0$.

Let now $(\Si, M, \pi, \SO(\eta))$ be a principal bundle over the manifold $M$ and
let $(x^\mu, S^a_b)$ be (overdetermined) fibered  ``coordinates'' on the principal bundle $\Si$.
We can define a right-invariant  pointwise basis $\si_{ab}$ for vertical vectors on $\Si$ by setting
$$
\si_{ab}= \eta_{d[a} \rho^d_{b]}
\qquad\qquad
\rho^d_{b}= S^d_c {\del\over\del S^{b}_c}
\fn$$
where $\eta_{ab}$ is the canonical diagonal matrix of signature $\eta=(r,s)$ and square brackets denote skew-symmetrization over indices.

A connection on $\Si$ is in the form
$$
\om= dx^\mu \otimes \(\del_\mu -\om^{ab}_{\mu} \si_{ab}\)
\fn$$
Also in this case the connection on $\Si$ induces connections on any associated bundle and there defines covariant derivatives of sections.

A {\sl frame\/} is a bundle  map $e:\Si\arr L(M)$ which preserves the right action, i.e.~such that
$$
\begindc{\commdiag}[1]
\obj(0,80)[Si]{$\Si$}
\obj(80,80)[LM]{$L(M)$}
\obj(0,20)[M1]{$M$}
\obj(80,20)[M2]{$M$}
\mor{Si}{M1}{}
\mor{LM}{M2}{}
\mor{Si}{LM}{$e$}
\mor{M1}{M2}{}[\atleft, \solidline] \mor(0,23)(80,23){}[\atleft, \solidline]
\enddc
\qquad\qquad\qquad
\begindc{\commdiag}[1]
\obj(0,80)[Si]{$\Si$}
\obj(80,80)[LM]{$L(M)$}
\obj(0,20)[Si1]{$\Si$}
\obj(80,20)[LM1]{$L(M)$}
\mor{Si}{Si1}{$R_S$}
\mor{LM}{LM1}{$R_{i(S)}$}
\mor{Si}{LM}{$e$}
\mor{Si1}{LM1}{$e$}
\enddc
\fn$$
i.e.~$e \circ R_S = R_{i(S)} \circ e$, where $R$ denotes the relevant canonical right actions
defined on both principal bundles $\Si$ and $L(M)$ and where $i:\SO(\eta)\arr \GL(m)$ is the
canonical group inclusion.  We stress that on any $M$ which allows global metrics of signature
$\eta$ the bundle $\Si$ can always be chosen so that there exist global frames; see \ref{Jadwisin}.
Locally the frame is represented by invertible matrices $e_a^\mu$ and it defines a spacetime metric
$g_{\mu\nu}= e_a^\mu \>\eta_{ab}\> e_b^\nu$ which is called the {\sl induced metric}.

As for the Levi-Civita connection, a frame defines a connection on $\Si$ (called the spin-connection of the frame) given by
$$
\om^{ab}_\mu= e^a_\al \( \Ga^\al_{\be\mu} e^{b\be} + d_\mu e^{b\al}\)
\fl{FrameConnection}$$
where $\{g\}^\al_{\be\mu}$  denote Christoffel symbols of the induced metric.
The spin-connection is compatible with the frame in the sense that
$$
\na_\mu e^\nu_a = d_\mu e^\nu_a + \{g\}^\nu_{\la\mu} e^\la_a - \om^c{}_{a\mu} e^\nu_c \equiv 0
\fn$$

{
\bib{Godina}{L. Fatibene, M. Ferraris, M. Francaviglia, M. Godina,
%{\it A geometric definition of Lie derivative for Spinor Fields},
in: Proceedings of {\sl ``6th International Conference on Differential Geometry
and its Applications, August 28--September 1, 1995"}, (Brno, Czech Republic),
Editor: I. Kol{\'a}{\v r}, MU University, Brno, Czech Republic (1996) 549.}
}

In general the (natural) lift $\hat \xi$ to $L(M)$ of a spacetime vector field $\xi$ is not adapted
to the image $e(\Si)\subset L(M)$ and thence it does not define a vector field on $\Si$.
%However, one can use shewsymmetrization of tetrad indices to define an adapted vector field.
With this notation the Kosmann lift of $\xi=\xi^\mu\del_\mu$  is defined by $\hat\xi_K= \xi^\mu\del_\mu + \hat \xi^{ab}\si_{ab}$ (see \ref{Godina}) where we set:
$$
\hat \xi^{ab}=  e_\nu^{[a} \na_\mu\xi^\nu   e^{b]\mu} -\om^{ab}_\mu \xi^\mu
\fl{KosmannLift}$$ 
and where $e^{a\mu}=\eta^{ac}e_c^\mu$ while $e_\nu^b$ denote the inverse frame matrix.

Let us stress that despite appearing so, the Kosmann lift \ShowLabel{KosmannLift} does not in fact depend on the connection, but just on the frame and its first derivatives. The same lift can be written as $\hat \xi^{ab}=  \na^{[b}\xi^{a]}   -\om^{ab}_\mu \xi^\mu$ where we set $\xi^a=\xi^\mu e^a_\mu$
since one can prove that
$$
\na_b \xi^a=e_\nu^{a} \na_\mu\xi^\nu   e^{\mu}_b
\fn$$

Another useful equivalent expression for the Kosmann lift is giving the vertical part of the lift with respect to the spin connection (see \ref{Libro}, pages 288--290), namely
$$
\hat \xi^{ab}_{(V)}:= \hat \xi^{ab} +  \om^{ab}_\mu \xi^\mu =  e_\nu^{[a} \na_\mu\xi^\nu   e^{b]\mu}= \na^{[b} \xi^{a]}
\fl{KLV}$$
This last expression is useful since it expresses a manifestly covariant quantity.

We have to stress that the Kosmann lift does not preserve commutators. In fact if one  considers two spacetime vectors $\xi$ and $\ze$ and 
computes the Kosmann lift of the commutator $[\xi, \ze]$ one can easily prove that
$$
[\xi, \ze]\>{}\hat{}_{K}= [\hat \xi_K, \hat \ze_K] + \frac[1/2] e^a_\al \Lie_\ze g^{\al\la} \Lie_\xi g_{\la\be} e^{b\be} \si_{ab}
\fn$$
Thence only if one restricts to Killing vectors (i.e.~$\Lie_\xi g=0$) one recovers that the lift preserves commutators.

Let us now consider tetrad-affine formulation of GR: the fundamental fields are a Lorentz connection $\Ga^{ab}_\mu$ and a vielbein $e^a=e^a_\mu\>dx^\mu$.
The connection defines the curvature form $R^{ab}= \frac[1/2] R^{ab}{}_{\mu\nu} \>dx^\mu\land dx^\nu$.
Let us also set $e=\det |e^a_\mu|$, $R^a{}_\mu= R^{ab}{}_{\mu\nu} e_b^\nu$ and $R= R^{ab}{}_{\mu\nu} e_a^\mu e_b^\nu$;
here $e_b^\nu$ denotes the inverse frame matrix of  $e^b_\nu$.
The frame also defines a metric $g_{\mu\nu}=  e^a_\mu \eta_{ab}  e^b_\nu$ which in turn defines its Levi-Civita spacetime connection $\Ga^\al_{\be\mu}$.

On a spacetime of dimension $4$ let us consider the Lagrangian
$$
L_{tA}= R^{ab}\land e^c \land e^d\>\ep_{abcd}
\fn$$
By variation we obtain
$$
\eqalign{
 \de L_{tA}=&-2e e^\si_a\( R^a{}_\mu -\frac[1/2] R e^a_\mu \) e_d^\mu \>\de e^d_\mu
             - ep_{abcd}\na_\mu\(e^c_\rho e^d_\si\)\ep^{\mu\nu\rho\si} \de\Ga^{ab}_\mu\cr
            &+\ep_{abcd}\na_\mu\(e^c_\rho e^d_\si \de\Ga^{ab}_\mu\)\ep^{\mu\nu\rho\si}}
\fn$$
Thus one obtains field equations in the form
$$
\cases{
&R^a{}_\mu -\frac[1/2] R e^a_\mu=0\cr
&\na_{[\mu}\(e^{[c}_\rho e^{d]}_{\si]}\)=0\cr
}\fn$$
The second field equation forces the connection to be the connection induced by the frame,
i.e.~$\Ga^{ab}_\mu=\om^{ab}_\mu$; then the first equation forces the induced metric to obey Einstein
equations.

This Field Theory is dynamically equivalent to standard GR, in the sense that it obeys equivalent
field equations.  However, the theory is in fact richer in its physical interpretation, since the
use of different variables and action principles generate larger symmetry and extra conservation
laws.  In fact, this theory has a bigger symmetry group being both generally covariant and Lorentz
covariant.

Noether theorem implies then conservation of the current 
$$
\calE^\mu= 4e e_a^\mu e_b^\nu \Lie_\Xi \Ga^{ab}_\nu -\xi^\mu L_{tA}
\fn$$
along any Lorentz gauge generator $\Xi= \xi^\mu\del_\mu + \xi^{ab} \si_{ab}$.
The Lie derivative of a connection is given by
$$
\Lie_\Xi \Ga^{ab}_\nu= \xi^\la R^{ab}{}_{\la\nu} + \na_\nu \hat\xi^{ab}
\fn$$
where we set $\hat\xi^{ab}=\xi^{ab}+\xi^\la\Ga^{ab}_\la$.
 
Hence one obtains
$$
\eqalign{
\calE^\mu=&  4e e_a^\mu \(R^a{}_\mu-\frac[1/2]R e^a_\mu\) \xi^\la -  4\na_\nu\(e e_a^\mu e_b^\nu\)\hat\xi^{ab} + 4 \na_\nu\(e e_a^\mu e_b^\nu\hat\xi^{ab} \) 
}
\fn$$
The first and second terms vanish on-shell; hence one obtains
$$
\calE^\mu= 4 \na_\nu\(e e_a^\mu e_b^\nu\hat\xi^{ab} \) 
\fl{FreeNoetherCurrent}$$
Let us stress that this current depends only on the Lorentz generator $\hat\xi^{ab}$.

Here is the issue with physical interpretation: we have two equivalent formulations of Einstein GR.
Noether currents in one case depend on spacetime vector fields, while in tetrad-affine formulation
Noether currents depend on Lorentz generators that {\it a priori\/} have nothing to do with
spacetime transformations.  Let us stress of course that unless the spacetime is Minkowski, there is
no class of (global) spacetime diffeomorphisms representing {\sl Lorentz transformations}.

Considering the dynamical equivalence at the level of field equations and solution space, one would
like this equivalence to be extended at the level of conservation laws.  Moreover, some of the
conserved quantities in standard GR are known to be related to physical quantities such as energy,
momentum and angular momentum, while one would wish to be able to identify the corresponding
quantities  also in the second formulation.  Kosmann lift is in fact essential to relate Lorentz
generators to spacetime diffeomorphisms and the corresponding conservation laws.

The Noether current \ShowLabel{FreeNoetherCurrent} can be restricted by setting $\Xi=\hat \xi_K$, so
that one obtains
$$
\calE^\mu_{tA}= 4 \na_\nu\(e \na^\mu \xi^\nu  \) 
\fn$$
which corresponds to the standard conserved quantity associated to spacetime diffeomorphisms in GR
written in terms of Komar superpotential.  This (and only this) restores the equivalence between
standard GR and tetrad-affine formulation at the level of conservation laws.

As a further example let us consider the covariant Lagrangian:
$$
L_H= L_{tA} + \be R^{ab}\land e_a\land e_b 
\fn$$
which is known as Holst's Lagrangian.

By variations one obtains equations
$$
\cases{
&e^\mu_d\(R^a_\mu -\frac[1/2] R e^a_\mu \) e_a^\si- \be R_{d\rho\mu\nu} \ep^{\mu\nu\rho\si}=0\cr
&\na_{[\mu}\(e^{[c}_\rho e^{d]}_{\si]}\)=0\cr
}
\fn$$
The second equation still imposes $\{g\}^{ab}_\mu=\om^{ab}_\mu$; this in turns implies
$R^a{}_{[\rho\mu\nu]}=0$ (first Bianchi identity) and hence Einstein equations.  This shows how
Holst's Lagrangian provides an equivalent formulation of standard GR as well.

It is interesting to check if also in this case the equivalence is preserved also at level of conservation laws.
The Noether current is
$$
\calE_{H}^\mu= 4e e_a^\mu e_b^\nu \Lie_\Xi \Ga^{ab}_\nu  + e e_c^\mu e_d^\nu \ep^{cd}\_a\_b \Lie_\Xi \Ga^{ab}_\nu- \xi^\mu L_{H}
\fn$$ 
As in the previous case this can be recasted modulo terms vanishing on-shell as follows
$$
\calE_{H}^\mu- \calE_{tA}^\mu = \na_\nu\(e e_c^\mu e_d^\nu \ep^{cd}\_a\_b \hat \xi^{ab}\)
\fn$$
Again this has nothing to do with sacetime symmetries and in general it would affect conserved
quantities.  When Kosmann lift is again inserted into these conservation laws one obtains
$$
 \calE_{H}^\mu- \calE_{tA}^\mu = \na_\nu\( \na^\rho\xi^\si \ep^{\mu\nu}{}_{\rho\si} \)
\fn$$
that vanishes being the divergence of a divergence.  Hence once again the correspondence at the
level of conservation laws is preserved when the Kosmann lift is used.

\NewSection{Conclusions and Perspectives}

The kinematics and dynamics of most fundamental interaction Physics can be defined purely in
terms of symmetries.  Noether theorem and consevation laws define a considerable part of its
physical interpretation.

The role of strong conservation laws is still unclear.  One can imagine some role in the Hamiltonian
framework that is however, still unclear and it deserves further investigations.
 
From a fundamental viewpoint one should also investigate whether one could describe Nature in terms 
of the whole set of conservation laws, without selecting special conservation laws to be endowed with a special meanings.

\Acknowledgements

The authors wish to thank C.~Rovelli for discussions.

This work is partially supported by MIUR: PRIN 2005 on {\it Leggi di conservazione e termodinamica
in meccanica dei continui e teorie di campo}.  We also acknowledge the contribution of INFN
(Iniziativa Specifica NA12) and the local research funds of Dipartimento di Matematica of Torino
University.

This paper is published despite the effects of the Italian law 133/08.  This law drastically reduces
public funds to public Italian universities, which is particularly dangerous for free scientific
research, and it will prevent young researchers from getting a position, either temporary or
tenured, in Italy ({\tt http://groups.google.it/group/scienceaction}).  The authors are protesting
against this law to obtain its cancellation.

\ShowBiblio

\end